\documentclass[11pt]{article}

\usepackage{graphicx}                  % Include figure files
\usepackage{dcolumn}                   % Align table columns on decimal point
\usepackage{bm}

\usepackage{jheppub}                   % jhep style

\makeatletter
\gdef\@fpheader{\ }                    % hack the jhep header
\makeatother

\usepackage{amsmath,amssymb}           % amstex
\usepackage{amscd}                     % ams commutative diagrams
\usepackage{epsfig}                    % ps figures
\usepackage[matrix,arrow]{xy}          % commutative diagrams
\usepackage{xspace}                    % space in abbreviations
\usepackage{stmaryrd}                  % extra symbol font (dbl bracket)
\usepackage{slashed}
\usepackage{tikz}

\usepackage{float}
\restylefloat{table}

\DeclareSymbolFont{bbold}{U}{bbold}{m}{n}
\DeclareSymbolFontAlphabet{\mathbbold}{bbold}

\newcommand{\bq}{\begin{equation}}
\newcommand{\eq}{\end{equation}}
\newcommand{\bea}{\begin{eqnarray}}
\newcommand{\eea}{\end{eqnarray}}

\newcommand{\dd}{\mathrm{d}}
\newcommand{\ee}{\mathrm{e}}

\newcommand{\der}{\partial}

\newcommand{\bbZ}{\mathbb{Z}}
\newcommand{\bbR}{\mathbb{R}}

\DeclareMathOperator{\SU}{\mathit{SU}}
\DeclareMathOperator{\SO}{\mathit{SO}}
\DeclareMathOperator{\SL}{\mathit{SL}}
\DeclareMathOperator{\GL}{\mathit{GL}}
\DeclareMathOperator{\USp}{\mathit{USp}}
\DeclareMathOperator{\Symp}{\mathit{USp}}
\DeclareMathOperator{\Spin}{\mathit{Spin}}

\DeclareMathOperator{\Cliff}{Cliff}

\DeclareMathOperator{\re}{Re}
\DeclareMathOperator{\im}{Im}

\newcommand{\rep}[1]{\mathbf{#1}}
\newcommand{\repp}[2]{(\rep{#1}, \rep{#2})}

\newcommand{\Gs}[1]{\Gamma(#1)}

\newcommand{\BLie}[2]{\left[#1,#2\right]}

\newcommand{\Dgen}{{D}}

\DeclareMathOperator{\adj}{ad}

\newcommand{\LC}{\nabla}

\newcommand{\GenRic}{R^{\scriptscriptstyle 0}}

\newcommand{\GenS}{R}

\newcommand{\proj}[1]{\times_{#1}}

\DeclareMathOperator{\Edd}{\mathit{E_{d(d)}}}
\DeclareMathOperator{\Hd}{\mathit{H_d}}
\DeclareMathOperator{\HdL}{\mathit{H^{\text{L}}_d}}
\DeclareMathOperator{\dHd}{\mathit{\tilde{H}_d}}

\newcommand{\tF}{{\tilde{F}}}

\newcommand{\am}{Q}

\newcommand{\Mint}{M}

\newcommand{\Kgen}{K}
\newcommand{\HConSp}{\Kgen_{\SU(8)}}
\newcommand{\GConSp}{\Kgen_G}	

\newcommand{\Tgen}{W}		 	
\newcommand{\Tint}{\Tgen_{\text{int}}}
\newcommand{\TTint}{T_{\text{int}}}
\newcommand{\HTConSp}{\Tgen_{\SU(8)}}	
\newcommand{\GTConSp}{\Tgen_G}	

\newcommand{\Ugen}{U}
\newcommand{\HUConSp}{\Ugen_{\SU(8)}}
\newcommand{\GUConSp}{\Ugen_G}

\newcommand{\CC}{\text{c.c.}}

\title{Supersymmetric Backgrounds and Generalised Special Holonomy}

\author[a]{Andr\'e Coimbra,}
\emailAdd{andre.coimbra@itp.uni-hannover.de}
\author[b,c]{Charles Strickland-Constable}
\emailAdd{charles.strickland.constable@desy.de}
\author[c,d]{and Daniel Waldram}
\emailAdd{d.waldram@imperial.ac.uk}

\affiliation[a]{Institut f\"ur Theoretische Physik \& 
  Center for Quantum Engineering and Spacetime Research,\\
  Leibniz Universit\"at Hannover, Appelstra{\ss}e 2, 
  30167 Hannover, Germany }

\affiliation[b]{II. Institut f\"ur Theoretische Physik
   der Universit\"at Hamburg, \\
   Luruper Chaussee 149, D-22761 Hamburg, Germany}

\affiliation[c]{Department of Physics,
   Imperial College London, \\
   Prince Consort Road, London, SW7 2AZ, UK}

\affiliation[d]{Berkeley Center for Theoretical Physics,
   LeConte Hall MC 7300, \\
   University of California, Berkeley, CA 94720, U.S.A.}

\subheader{\textrm{UCB-PTH-14/39 \\
      Imperial/TP/14/DW/04 \\ 
      ITP-UH-20/14}}

\abstract{We define intrinsic torsion in generalised geometry and use it to introduce a new notion of generalised special holonomy. We then consider generic warped supersymmetric flux compactifications of M theory and Type II of the form $\mathbb{R}^{D-1,1}\times M$. Using the language of $\Edd\times\bbR^+$ generalised geometry, we show that, for $D\geq 4$, preserving minimal supersymmetry is equivalent to the manifold $M$ having generalised special holonomy and list the relevant holonomy groups. We conjecture that this result extends to backgrounds preserving any number of supersymmetries. As a prime example, we consider $\mathcal{N}=1$ in $D=4$. The corresponding generalised special holonomy group is $\SU(7)$, giving the natural M theory extension to the notion of a $G_2$ manifold, and, for Type II backgrounds, reformulating the pure spinor $\SU(3)\times\SU(3)$ conditions as an integrable structure. 
}

\begin{document}   
\maketitle

%%%%%%%%%%%%%%%%%%%%%%%%%%%%%%%%%%%%%%%%%%%%%%%%%%%%%%%%%%%%%%%%%%%%%%%%%%

\section{Introduction}
\label{sec:intro}

Understanding the geometry of supersymmetric flux backgrounds of string and M theory is a central question both for string phenomenology and for the study of gauge-gravity duality. Without the presence of fluxes, it is well known that supersymmetry implies that the background is a special holonomy manifold. The possible special holonomy groups $G$ have been classified~\cite{Berger}, and include, for example the classic cases of Calabi--Yau spaces with $\SU(n)$-holonomy and seven-dimensional Joyce manifolds with $G_2$-holonomy. Special holonomy is equivalent to the existence of an integrable or ``torsion-free'' $G$-structure and this integrability is central to  understanding their properties, such as their moduli and Ricci-flatness. 

For flux backgrounds the situation becomes more complicated. The manifolds no longer have special holonomy. However, since the work of~\cite{Strominger:1986uh,chris-het}, it has been known that certain cases could still be described in terms of known geometrical structures. In general, the Killing spinors still locally define a natural $G$-structure, and the application of the $G$-structure formalism, following~\cite{Gauntlett:2002sc,Gauntlett:2002nw,Gauntlett:2002fz}, has become a standard tool for analysing the geometry. The lack of integrability is characterised by the ``intrinsic torsion'' of the $G$-structure, and the Killing spinor conditions can be interpreted as setting some components of the intrinsic torsion to zero, while relating the remaining components to the fluxes. 

In this paper, we will reinterpret generic supersymmetric flux backgrounds of M theory and Type II string theory as \emph{integrable, globally defined structures}, but in generalised rather than conventional geometry. In particular, we argue that when compactifying to flat space in four or more dimensions, the supersymmetry conditions are equivalent to saying the internal manifold has special holonomy in a generalised sense. Rephrasing supersymmetry as a integrability condition is useful since it gives new tools for addressing questions such the moduli space of backgrounds, and for building new explicit solutions. We do not address these questions here, except to give give a short proof that, in analogy to the conventional case, these generalised holonomy manifolds are also generalised Ricci-flat. Physically this is equivalent to showing that, given the Bianchi identities for the fluxes, supersymmetry implies we have a solution of the equations of
motion~\cite{Gauntlett:2002fz,Gauntlett:2002nw,Gauntlett:2002sc,Gauntlett:2003cy,Koerber:2007hd}.

The idea of generalised integrable structures, in particular generalised complex structures, was central in the original $O(d)\times O(d)$ version of generalised geometry, due to Hitchin and Gualtieri~\cite{GCY,Gualtieri}, which ``geometrises'' the NSNS sector of the fluxes. These structures were first applied to string $\sigma$-models in~\cite{sigmas}. The connection to supergravity backgrounds appeared in~\cite{GMPT}, where it was shown that, in the absence of RR fluxes, the Killing spinor equations amount to an integrability condition on a generalised $G$-structure, specifically that the internal space is a six-dimensional ``generalised Calabi--Yau metric'' manifold, admitting a pair of generalised complex structures defining an $\SU(3)\times\SU(3)$ generalised structure group. (For the seven-dimensional case one obtains instead $G_2\times G_2$ manifolds~\cite{Jeschek:2004wy,Witt:2004vr}.) The integrability is again then partially violated by the presence of RR fluxes. For supersymmetric backgrounds, among many other topics, these methods were applied to other dimensions in~\cite{Lust:2010by}, they have provided tools for the investigation of new background solutions (see for example~\cite{Grana:2006kf,Andriot:2008va} and more recently \cite{Apruzzi:2013yva,Apruzzi:2014qva}) as well as the AdS/CFT correspondence~\cite{Minasian:2006hv,Butti:2007aq,Gabella:2009wu}, and have recently also been applied to the analysis of full ten-dimensional backgrounds in~\cite{Tomasiello:2011eb}. For a much more thorough review of some of these and other aspects of flux compactifications in string theory, see~\cite{Grana:2005jc}.

In addition, a notion of generalised holonomy has previously been considered by several authors~\cite{Duff:1990xz,Berkooz:1996km,FigueroaO'Farrill:2002ft,gen-holo-duff,Hull:2003mf,Lu:2005im}. However, in these works the holonomy being considered is that of a conventional (Clifford-algebra-valued) connection that appears explicitly in the Killing spinor equations. By contrast, here we consider \emph{generalised connections}, of which only certain covariant projections, analogues of Dirac operators, appear in the Killing spinor equations. It is precisely the fact that we have generalised derivatives that will allow us to define the torsion of the connection and hence formulate a new notion of integrability which determines supersymmetric backgrounds. One should also note that, here, the term ``generalised special holonomy'' is somewhat misleading since, in contrast to~\cite{Duff:1990xz,Berkooz:1996km,FigueroaO'Farrill:2002ft,gen-holo-duff,Hull:2003mf,Lu:2005im}, the definition actually follows from a notion of integrability, rather than of generalised parallel transport and holonomy. In conventional geometry the two are equivalent. It is an interesting open question to define an appropriate notion of holonomy for generalised connections and to show that here too the two concepts agree.  

The concept of an integrable generalised $G$-structure can be defined for any version of generalised geometry. They include for instance generalised complex structures and generalised Calabi--Yau geometries in the case of $O(d,d)\times\bbR^+$ generalised geometry.  However, here we will specifically focus on $\Edd\times\bbR^+$ generalised geometry, also known as extended or exceptional generalised geometry~\cite{chris,PW}, as developed in~\cite{CSW2,CSW3}. This gives a unified geometric description applicable to both M theory and Type II theories, restricted to $d$ or $(d-1)$ dimensions respectively, accommodating all fluxes, including the RR fields in Type II. The bosonic fields combine into a generalised metric, invariant under local $\Hd\subset\Edd\times\bbR^+$ changes of frame, and one can define an analogue of the Levi--Civita connection, a generalised torsion-free $\Hd$ connection, that determines both the bosonic and fermionic dynamics. Closely related work, all with a putative enlarged spacetime, includes the M-theory discussion of~\cite{BP} (a cousin of Double Field Theory~\cite{dft}) and the recent extensions to a full ten- or eleven-dimensional theory given in~\cite{EFT1a,EFT1b,EFT2}. That the language of $\Edd\times\bbR^+$ generalised geometry is particularly well suited to describe supersymmetric compactifications has already been shown in~\cite{GLSW,GO,GO2} (we will comment on the relation between the last two works and ours in the conclusion). The advantage of $\Edd\times\bbR^+$ generalised geometry is that it encompasses the different classes of ordinary $G$-structures that appear in supersymmetric backgrounds as a single global generalised structure group $G\subset\Hd$, even in the presence of RR flux. 

For any generalised geometry it is natural to define special holonomy manifolds as those admitting a torsion-free $G$-structure, where $G$ is a proper subgroup of the group $\Hd$ that preserves the generalised metric. In this paper we show that, for warped  compactifications of the form $\bbR^{D-1,1}\times M$ with $D\geq 4$, minimally supersymmetric backgrounds are in one-to-one correspondence with $M$ having generalised special holonomy, and list the corresponding holonomy groups. We also conjecture that the equivalence of supersymmetry and special holonomy extends to backgrounds preserving any number of supersymmetries. 

As a standard example, we consider in some detail $D=4$, where $\mathcal{N}=1$ supersymmetry implies $M$ has $\SU(7)$ generalised special holonomy. This gives a natural M theory extension of a conventional $G_2$ manifold. As a Type II background it provides an integrable reformulation of the pure spinor $\SU(3)\times\SU(3)$ structures discussed in~\cite{GMPT}. 

Note that for all warped flux compactifications to Minkowski backgrounds there are standard ``no-go'' theorems~\cite{Candelas,Maldacena:2000mw,Ivanov:2000fg,Giddings:2001yu,Gauntlett:2002sc,Gauntlett:2003cy} that exclude the possibility of compact $M$ in the absence of sources. Thus, the generic backgrounds described in this paper, should be viewed either as non-compact, or as manifolds with boundaries where the sources have been excised. We also do not consider backgrounds with $D\leq 3$ since, as is explained there, the construction of~\cite{CSW2} breaks down for $E_{8(8)}$ and larger exceptional groups (though see also~\cite{Berman:2012vc,EFT1b,Rosabal:2014rga}). We also stay in $D\leq 7$ since beyond that exceptional generalised geometry does not add much to the conventional description~\cite{CSW2,CSW3}. Indeed, as we will remark in section~\ref{sec:genD}, already at $D=7$ the results we present in this paper reduce to the conventional picture.

The paper is organised as follows. We begin in section~\ref{sec:susyback} by establishing our conventions for compactifications to $D$-dimensional flat space and then reviewing the conventional $G$-structure analysis and some of the properties of our main $\mathcal{N}=1$, $D=4$ example. In section~\ref{sec:gg} we give a very quick review of $\Edd\times\bbR^+$ generalised geometry, introduce the concept of generalised intrinsic torsion and special holonomy, and give a simple example of a generalised special holonomy manifold, the intersection of three M5 branes. We prove our main result in the case $D=4$ in section~\ref{sec:genbackgrounds}, showing that $\mathcal{N}=1$ backgrounds are in one-to-one correspondence with manifolds of $\SU(7)$ special holonomy. In section~\ref{sec:genD} we extend this result to all $D\geq 4$ backgrounds preserving minimal supersymmetry. We conclude with some discussion in section~\ref{sec:conc}.

%%%%%%%%%%%%%%%%%%%%%%%%%%%%%%%%%%%%%%%

\section{Supersymmetric Minkowksi backgrounds}
\label{sec:susyback}

In this paper we consider generic supersymmetric flux compactifications of M theory and Type II string theory to four- and higher-dimensional  Minkowski space. This means we have the warped metric ansatz
\begin{equation}
\label{eq:g}
   \dd s^2 = \ee^{2A}\dd s^2(\bbR^{D-1,1}) + \dd s^2(M) . 
\end{equation}
with $D\geq 4$ and where the warp factor $A$ is a scalar function of the internal coordinates. The internal space $M$ is a spin manifold with Riemannian metric $g$, of dimension $d$ in M theory and $d-1$ in Type II. To match the conventions of~\cite{CSW2,CSW3}, we take $A=\Delta$ in M theory, so that $A=\Delta+\frac{1}{3}\phi$ in Type II, where $\phi$ is the dilaton, and the metric is in the string frame. For the fluxes we keep only the components consistent with the $D$-dimensional Lorentz symmetry. 

Focussing on the M theory case, this means that of the eleven-dimensional four-form flux $\mathcal{F}$  we keep only the components that can be viewed as internal four- and seven-form fluxes $F$ and $\tilde{F}$. Using $m,n,\dots = 1, \dots , d$ for internal coordinate indices, these are given by 
\begin{equation*}
\begin{aligned}
  &F_{m_1 \dots m_4} = \mathcal{F}_{m_1 \dots m_4} ,\\  
  &\tF_{m_1 \dots m_7} 
       = \left(*\mathcal{F}\right)_{m_1 \dots m_7} .
\end{aligned} 
\end{equation*}

If $\varepsilon$ is an eleven-dimensional spinor, the conditions for a supersymmetric background are the Killing spinor equations, which can be written as\footnote{Throughout this paper we follow the conventions of~\cite{CSW2,CSW3}}
\begin{equation}
\label{eq:susy-ferm}
\begin{split}
   \left[ 
      \Gamma^m\LC_m  - \tfrac{1}{96} \Gamma^{m_1 \dots m_4} F_{m_1 \dots m_4}    - \tfrac{1}{4} \tfrac{1}{7!} \Gamma^{m_1 \dots m_7} \tilde{F}_{m_1 \dots m_7} 
      + \tfrac{1}{2}(9-d) \Gamma^m(\der_m \Delta) 
      \right] \varepsilon &= 0 ,\\
   \left[ 
      \LC_m + \tfrac{1}{288} F_{n_1 \dots n_4} \left(
         \Gamma_m{}^{n_1 \dots n_4} 
         - 8 \delta_{m}{}^{n_1} \Gamma^{n_2 n_3 n_4}\right)  
      - \tfrac{1}{12} \tfrac{1}{6!} \tilde{F}_{mn_1 \dots n_6} 
	 \Gamma^{n_1 \dots n_6} 
      \right] \varepsilon &= 0 , 
\end{split}
\end{equation}
where $\Gamma^m$ are eleven-dimensional gamma matrices. Note also that if one traces the first equation with $\Gamma^m$ and subtracts it from the second, one obtains a purely algebraic equation relating the gauge fluxes to the warp factor, which is the more conventional way the conditions appear. 

Our basic example in this paper will be the case of $d=7$ preserving $\mathcal{N}=1$ supersymmetry in four-dimensions~\cite{Kaste:2003zd,Behrndt:2003zg,Dall'Agata:2003ir,Lukas:2004ip}. The eleven-dimensional spinors $\varepsilon$ are decomposed into four- and seven-dimensional spinors $\eta^{\pm}$ and $\epsilon$ respectively according to
\begin{equation}
\label{eq:mtheoryspinor}
   \varepsilon = \eta^+ \otimes \epsilon + \eta^- \otimes \epsilon^* .
\end{equation}
where $\pm$ denotes the chirality of the four-dimensional spinor. Here the internal spinor $\epsilon$ is complex, and can be thought of as a pair of real $\Spin(7)$ spinors $\epsilon = \re\epsilon + i \im\epsilon$. If one has an $\mathcal{N}=1$ vacuum, then there exists a single spinor field $\epsilon$ which satisfies the Killing spinor equations
\begin{equation}
\label{eq:elevendsusy}
\begin{split}
   \LC_m \epsilon + \tfrac{1}{288} 
   (\gamma_m{}^{n_1 \dots n_4} - 8 \delta_{m}{}^{n_1} \gamma^{n_2 n_3 n_4}) 
   F_{n_1 \dots n_4} \epsilon- \tfrac{1}{12} \tfrac{1}{6!} \tilde{F}_{mn_1 \dots n_6} 			\gamma^{n_1 \dots n_6} \epsilon &=0,\\[8pt]
\gamma^m\LC_m \epsilon + \gamma^m(\der_m \Delta) \epsilon - \tfrac{1}{96} \gamma^{m_1 \dots m_4} F_{m_1 \dots m_4} \epsilon - \tfrac{1}{4} \tfrac{1}{7!} \gamma^{m_1 \dots m_7} \tilde{F}_{m_1 \dots m_7}  \epsilon &=0,
\end{split}
\end{equation}
where $\LC$ is the Levi--Civita connection for $g$ and $\gamma^m$ are the $\Cliff(7;\bbR)$ gamma matrices. The Killing spinor equations imply that $\tF$, which can only be present for $d=7$, in fact vanishes for Minkowski backgrounds~\cite{Kaste:2003zd}, since it can only be supported by a cosmological constant. They also imply that
\begin{equation}
\label{eq:norms}
   \bar{\epsilon}\epsilon = \text{const} \times \ee^{\Delta} , \qquad 
   \epsilon^T\epsilon = \text{const} \times \ee^{-2\Delta} ,
\end{equation}
so that the rescaled Killing spinor $\ee^{-\Delta/2} \epsilon$ has constant norm. By phase rescaling of $\eta^\pm$ one can also set $\epsilon^T\epsilon$ to be real. In this case, the individual $\Spin(7)$ spinors $\re\epsilon$ and $\im\epsilon$ are then orthogonal, but their norms need not be constant~\cite{Lukas:2004ip}.

Although we do not give the details, similar expressions for the Killing spinor equations exist in the Type II case and were analysed in~\cite{GMPT}. In this case, there are a pair of real ten-dimensional spinors $(\varepsilon_1,\varepsilon_2)$ which decompose under $\Spin(3,1)\times\Spin(6)$ as   
\begin{equation}
\begin{aligned}
   \varepsilon_1 = \eta^+\otimes \epsilon^+_1 
       +  \eta^-\otimes \epsilon^-_1 , \\
   \varepsilon_2 = \eta^+\otimes \epsilon^\mp_2 
       +  \eta^-\otimes \epsilon^\pm_2 , 
\end{aligned}
\end{equation}
where the $\pm$ superscripts denote chiralities, $\epsilon_i^-$ and $\eta^-$ are the charge conjugates of $\epsilon_i^+$ and $\eta^+$ respectively, and the two choices of signs in the second line refer to Type IIA and IIB. The two internal spinors can be combined into a single, complex, eight-component object
\begin{equation}
\label{eq:IIspinor}
   \epsilon = \ee^{-\phi/6}\begin{pmatrix} \epsilon^+_1 \\ \epsilon^-_2 \end{pmatrix} ,
\end{equation}
which for Type IIA is the standard lift to the $d=7$ complex spinor of the M theory reduction. Supersymmetry again implies that~\cite{GMPT} 
\begin{equation}
\label{eq:normsII}
   \bar{\epsilon}\epsilon 
      = \ee^{-\phi/3}\left(\bar{\epsilon}^+_1\epsilon_1^+ 
          + \bar{\epsilon}^+_2\epsilon^+_2 \right) 
      = \text{const.}\times\ee^\Delta , 
\end{equation}
in both Type IIA and Type IIB. However, the individual norms of $\epsilon_i^+$ are not necessarily constant. 

\subsection{Supersymmetry and $G$-structures}
\label{sec:Gstruct}

In the past, the standard and very productive way to study supersymmetric flux backgrounds has been using the machinery of $G$-structures~\cite{Gauntlett:2002sc,Gauntlett:2002nw,Gauntlett:2002fz}. We will quickly review some of the key ideas of this approach, again focussing on the $D=4$, $\mathcal{N}=1$ example following~\cite{Kaste:2003zd,Lukas:2004ip} and~\cite{GMPT}. Some further definitions of $G$-structures and intrinsic torsion are given in appendix~\ref{app:Tint}.

In the case of vanishing fluxes, the Killing spinor equations~\eqref{eq:susy-ferm} are equivalent to special holonomy, as they become $\LC_m \varepsilon = 0$ with $\LC$ the Levi--Civita connection on $\Mint$. In the $D=4$ example in M theory, restricting to $\mathcal{N}=1$, one can always choose $\epsilon$ to be real. This gives a single covariantly constant spinor on $\Mint$, and hence implies the manifold has $G_2$ holonomy. For Type II, a covariantly constant spinor in six dimensions defines a Calabi--Yau structure. Since both $\epsilon_1^+$ and $\epsilon_2^+$ can be proportional to the constant spinor, this actually always leads to an $\mathcal{N}=2$ background. 

The inclusion of fluxes introduces extra complications to the picture. First, the stabliser subgroup of $\Spin(d)$ that leaves the Killing spinors invariant can vary over the manifold, and second, and most significantly, the integrability of the corresponding $G$-structures is spoiled since they are no longer covariantly constant with respect to the Levi--Civita connection. Nonetheless, one can neatly characterise the geometry of the Killing spinor equations as determining the intrinsic torsion of the $G$-structure in terms of the fluxes. 

The $d=7$ example exhibits all these features. Consider first the M theory case. The flux terms in~\eqref{eq:elevendsusy} mix the real and imaginary parts of the complex spinors so one cannot choose the spinor to be purely real. Thus, at generic points on the manifold, one has a pair of real $\Spin(7)$ spinors, and hence a pair of $G_2$ structures, or equivalently, an $\SU(3)$ structure. However, at some points, one of the spinors can potentially vanish, and the stabiliser group becomes simply $G_2$. Although the $\SU(3)$ structure is not necessarily global, its intrinsic torsion captures the supersymmetry conditions. One way to see this is to note that the $\SU(3)$ structure defines a set of invariant forms $(v,J,\Omega)$. The one-form $v$ defines a local product structure splitting $TM$ into the sum of one- and  six-dimensional bundles. The forms $J$ and $\Omega$ are a symplectic form and a complex three-form respectively defining a conventional $\SU(3)$ structure on the six-dimensional bundle. The intrinsic torsion of the $\SU(3)$ structure is then captured by the exterior derivatives of $(v,J,\Omega)$ (see appendix~\ref{app:Tint} for more discussion). For example, in the special case where $\epsilon^T\epsilon=0$, which implies the real spinors have equal constant norms, one has~\cite{Kaste:2003zd} 
\begin{equation}
\begin{aligned} 
   \dd \left( \ee^{2\Delta} v \right) &= 0 , & && &&
   \dd \left( \ee^{4\Delta} J \right) &=  - 4 \ee^{4\Delta} * F , & && &&
   \dd \left( \ee^{3\Delta} \Omega \right) &= 0 . 
\end{aligned}
\end{equation}
We see that some of the components of the intrinsic torsion vanish, and others are related to the flux $F$. More generally, one must also keep track of function $\epsilon^T\epsilon$ as well as the forms $(v,J,\Omega)$, the $\SU(3)$ structure is not global, and the intrinsic torsion conditions become considerably more complicated~\cite{Lukas:2004ip}. 

In the Type II case, the two $\Spin(6)$ spinors $\epsilon_i^+$ each define an $\SU(3)$ structure, so that together the structure is generically $\SU(2)$. However, there can be points on the manifold where the spinors become parallel, so that the stabiliser group enlarges to $\SU(3)$. Also one of the spinors can vanish, for example in solutions arising from NS fivebranes wrapped on K\"ahler two-cycles in a Calabi--Yau manifold~\cite{Strominger:1986uh,chris-het,Gauntlett:2001ur}, in which case the structure is $\SU(3)$. Again, although the $\SU(2)$ structure is not global, its intrinsic torsion captures the supersymmetry conditions. The remarkable reformulation of~\cite{GMPT} showed that the pair of $\SU(3)$ structures could actually be viewed as an $\SU(3)\times\SU(3)$ structure in $O(6,6)$ generalised geometry, even at the points where the spinors became parallel, though this fails to capture the case where one of the spinors vanishes. The invariant tensors are odd and even poly-forms $\Phi^\pm\in\Lambda^\pm T^*M$, and satisfy~\cite{GMPT,Grana:2006kf} (the upper and lower signs refer to Type IIA and IIB respectively)
\begin{equation}
\begin{aligned}
   \dd \left( \ee^{2A}\Phi^\pm \right) &= 0 , \\
   \dd \left( \ee^{A}\re\Phi^\mp \right) 
      &= \tfrac{1}{16}\ee^{2A}(|a|^2-|b|^2) F^\pm , \\
   \dd \left( \ee^{3A}\im\Phi^\mp \right) 
      &= \tfrac{1}{16}\ee^{2A}(|a|^2+|b|^2) *\lambda(F^{\pm}) , 
\end{aligned}
\end{equation}
where $\Phi^\pm$ incorporate the NSNS $B$-field and the dilaton, $F^\pm\in\Lambda^\pm T^*M$ is a polyform of RR fluxes, again with the $B$ field incorporated, $*\lambda$ is a natural $O(d,d)$ operation that dualises $F^\pm$, and $|a|^2$ and $|b|^2$ are the norms of the two spinors $\eta_i^+$.  The presence of the RR fluxes on the right-hand side of these equations, indicates that the generalised Calabi--Yau structure becomes non-integrable when they are present. 

Our goal in the following is to see how the extension to $\Edd\times\bbR^+$ generalised geometry, which geometrises all the flux degrees of freedom, provides a global definition of $G$-structure in all cases, and restores a notion of integrability even to generic flux backgrounds.

%%%%%%%%%%%%%%%%%%%

\section{Generalised geometry}
\label{sec:gg}

Let us briefly review the structure of $\Edd\times\bbR^+$ generalised geometry as first introduced in~\cite{CSW2,CSW3}. In generalised geometry one replaces the tangent bundle of the manifold $TM$ with a larger generalised tangent bundle $E$. Associated to $E$ there is also a generalised frame bundle $\tilde{F}$. Focussing on M theory, consider a $d$-dimensional spin manifold $M$ with $d\leq 7$. For $\Edd\times\bbR^+$ generalised geometry~\cite{chris,PW} one has 
\begin{equation}
   E \simeq TM \oplus \Lambda^2 T^*M \oplus \Lambda^5 T^*M  \oplus \left(T^*M \otimes \Lambda^7 T^*M\right) .
\end{equation} 
For the Type II theory, $M$ is a $(d-1)$-dimensional spin manifold and the generalised tangent space is~\cite{chris}
\begin{equation}
\label{eq:genEII}
   E \simeq TM \oplus T^*M \oplus \Lambda^\pm T^*M \oplus \Lambda^5T^*M 
       \oplus \left(T^*M\otimes\Lambda^6T^*M\right) ,
\end{equation}
where $\pm$ refers to Type IIA and IIB. In each case, the generalised frame bundle $\tilde{F}$ is an $\Edd\times\bbR^+$ principal bundle constructed from frames for $E$. The generalised tangent space is thus is an $\Edd\times\bbR^+$ vector bundle. For example, for $d=7$, it is associated to the $\rep{56_1}$ representation, where the subscript denotes the $\bbR^+$ weight. The case for general $d$ is summarised in table~\ref{tab:gg}. 

The differential structure of the generalised tangent bundle is described by the generalised Lie derivative $L_V V'\in\Gs{E}$, given two generalised vector fields $V, V' \in\Gs{E}$~\cite{CSW2} (also known as the Dorfman derivative or non-skew-symmetric Courant bracket). This gives $E$ the structure of a Leibniz algebroid~\cite{Baraglia:2011dg}. A generalised covariant derivative is a linear map $D:\Gs{E}\to\Gs{E^*\otimes E}$ which preserves the $E_{7(7)}\times\bbR^+$ structure and satisfies a Leibniz condition, that is, given a function $f$ and $V\in\Gs{E}$, then
\begin{equation}
   D(fV) = f(DV) + (\dd f)\otimes V ,
\end{equation}
where $\dd f$ is the one-form in $T^*M\subset E^*$. This property means one can extended the action of the derivative to any $\Edd\times\bbR^+$ vector bundle. We define the generalised torsion in complete analogy to ordinary geometry as 
\begin{equation}
T(D)(V,V')=L^D_V V' - L_V V',
\end{equation} 
where $L^D_V$ is the generalised Lie derivative calculated using the connection $D$. The generalised torsion will constrain a space $T(D)\in \Gs{\Tgen}$ where $W$ is a subbundle $\Tgen\subset E^* \otimes \text{ad}\tilde{F}$. The $\Edd\times\bbR^+$ representations of the $\Tgen$ vector bundles as given in table~\ref{tab:gg}. 
\begin{table}[htb]
\centering
\begin{tabular}{llll}
   $\Edd$ group & $\dHd$ group & $E$ & $\Tgen$ \\
   \hline
   $E_{7(7)}$ & $\SU(8)$ & $\rep{56}_\rep{1}$ 
      & $\rep{912}_\rep{-1} + \rep{56}_\rep{-1}$ \\
   $E_{6(6)}$ & $\USp(8)$ & $\rep{27}'_\rep{1}$  
      & $\rep{351}'_\rep{-1} + \rep{27}_\rep{-1}$ \\
   $E_{5(5)}\simeq\Spin(5,5)$ & $\USp(4)\times\USp(4)$
      & $\rep{16}^c_\rep{1}$
      & $\rep{144}^c_\rep{-1} + \rep{16}^c_\rep{-1}$ \\
   $E_{4(4)}\simeq\SL(5,\bbR)$ & $\USp(4)$ 
      & $\rep{10}'_\rep{1}$
      & $\rep{40}_\rep{-1} + \rep{15}'_\rep{-1} + \rep{10}_\rep{-1}$ 
\end{tabular}
\caption{Generalised tangent space and torsion representations}
\label{tab:gg}
\end{table}

A generalised metric $G$ defines an $\Hd$ sub-bundle $\tilde{P} \subset \tilde{F}$ of the frame bundle, where $\Hd$ is the maximal compact subgroup of $\Edd$. This is in complete analogy with a conventional metric $g$, which defines an $O(d)$ sub-bundle of the $\GL(d;\bbR)$ frame bundle for $TM$, and $G$ can similarly be viewed as an $E^*\otimes E^*$ generalised tensor which defines a positive definite norm $G(V,V)$ on the space of generalised vectors $V\in\Gs{E}$. Since the manifold is spin, one can introduce the double cover $\dHd$, under which spinors will transform~\cite{chris}. These groups are listed in table~\ref{tab:gg}. For $d=7$, $\tilde{H}_7$ is $\SU(8)$.  We say that a generalised connection is compatible with the generalised metric, or equivalently, is an $\Hd$-connection, if $D G = 0$. It turns out that unlike the case of Riemannian geometry, there exists a family of natural connections which are both metric compatible and torsion-free. All the dynamics and supersymmetry transformations of the theory are formulated in terms of this connection -- the bosonic equations correspond to a vanishing generalised Ricci tensor, and the fermion variations and equations of motion depend on two $\Hd$-covariant projections of $\Dgen$, the generalised analogues of Dirac operators. Remarkably, the Ricci tensor and the projected operators are uniquely determined, taking the same value independent of the choice of $\Dgen$. All these points are described in significantly more detail in~\cite{CSW2,CSW3}. 

We again stress that this formalism applies not only to M theory reductions but also to Type II theories with all fluxes, including RR, present. As was explained in~\cite{CSW2}, the two forms of the generalised tangent space~\eqref{eq:genEII} amount to reducing the dimension of the internal manifold by one and  picking an appropriate $\GL(d-1;\bbR)\subset \Edd$ substructure (the IIA and IIB theories correspond to two inequivalent embeddings). Previously, $\mathcal{N}=1$ and $\mathcal{N}=2$ compactifications of Type II theories with RR flux have been studied in exceptional generalised geometry in~\cite{GLSW, GO, GO2}, while~\cite{Lee:2014mla} used this language to describe Type IIB compactifications with maximal supersymmetry.

\subsection{Generalised intrinsic torsion}
\label{sec:TintGG}

The notion of generalised intrinsic torsion was discussed in a heterotic extension of $O(d,d)\times\bbR^+$ generalised geometry in~\cite{Coimbra:2014qaa}, and follows in complete analogy with the ordinary case (for a quick review of the latter see appendix~\ref{app:Tint}). Here we give a general definition. 

Let $E$ be the generalised tangent bundle and $\tilde{F}$ the corresponding generalised frame bundle. Let $\tilde{P}_G\subset\tilde{F}$ be a principal sub-bundle with group $G$. Let $\hat{\Dgen}$ be some generalised connection compatible with $\tilde{P}_G$. By definition, any other compatible connection $\hat{\Dgen}'$ can be written as $\hat{\Dgen}'=\hat{\Dgen}+\Sigma$ where 
\begin{equation}
   \Sigma = \hat{\Dgen} - \hat{\Dgen}' \in \Gs{\GConSp} , \qquad
   \text{with} \qquad \GConSp=E^*\otimes \adj{\tilde{P}_G} .
\end{equation}
We then define a map $\tau:\GConSp\to W$, where $W$ is the space of generalised torsions, as the difference of the torsions of $\hat{\Dgen}$ and $\hat{\Dgen}'$, 
\begin{equation}
   \tau(\Sigma) = T(\hat{\Dgen}) - T(\hat{\Dgen}') \in \Gs{W} .
\end{equation}
In general, the span of $\tau(\Sigma)$ may not fill out the whole of $W$. If we define the image 
\begin{equation}
   \GTConSp = \im \tau \subset \Tgen ,
\end{equation}
we can then simply define the space of the generalised intrinsic torsion, in exact analogy to ordinary geometry, as the part of $\Tgen$ not spanned by $\GTConSp$, that is 
\begin{equation}
   \Tint = \Tgen / \GTConSp.
\end{equation}
Given any $G$-compatible connection $\hat{\Dgen}$, we say that the generalised intrinsic torsion $\TTint(\tilde{P}_G)$, of the generalised $G$-structure $\tilde{P}_G$, is the projection of $T(\hat{\Dgen})$ onto $\Tint$. By definition this is independent of the choice of $\hat{\Dgen}$. It is the part of the torsion that cannot be changed by varying our choice of compatible connection. We can also use the kernel of the map $\tau$ to identify a subspace of connections $\GConSp$ as 
\begin{equation}
   \GUConSp =\ker \tau \subset \GConSp .
\end{equation}
Thus $\GUConSp$ is the space of compatible connections with a given fixed torsion.

The intrinsic torsion $\TTint(\tilde{P}_G)$ is the obstruction to finding a connection which is simultaneously torsion-free and compatible with the $G$-structure. If it vanishes we say that $\tilde{P}_G$ is an \emph{integrable} or \emph{torsion-free} $G$-structure. If, in addition, $G\subset\Hd$ where $\Hd$ is the maximally compact subgroup of the fibre group of the frame bundle $\tilde{F}$ (for $\Edd\times\bbR^+$ the $\Hd$ groups are listed in table~\ref{tab:gg}), we then say that $M$ is a manifold with $G$ \emph{generalised special holonomy}. This latter term is in some ways misleading since we have not defined the notions of parallel transport and holonomy for a generalised connection. Nonetheless, since for conventional geometry torsion-free $G$-structures on $M$ with $G\subset O(d)$ are equivalent to special holonomy, we use the same term here. 

When $G\subset\Hd$, the norm defined by the generalised metric $G$ provides unique decompositions 
\begin{equation}
\begin{aligned}
   \Tgen &= \GTConSp \oplus \Tint , \\
   \GConSp &= \GTConSp \oplus \GUConSp , 
\end{aligned}
\end{equation}
by taking the orthogonal complements of $\GTConSp$ in $\Tgen$ and $\GUConSp$ in $\GConSp$. 

As special case, consider $G=\Hd$. One of the main results proven in~\cite{CSW2} was that it is always possible to find torsion-free connections which are compatible with $\Hd$, but that the solution is not unique in general (unlike ordinary Riemannian geometry which singles out the Levi--Civita connection). This means $\Tint=0$ but $\GUConSp$ is non-trivial. To see this concretely, consider the $d=7$ example, where $H_7=\SU(8)/\bbZ_2$. The problem reduces to linear algebra at a point in the manifold, so we just need to know the representations of the corresponding vector bundles and, for the sake of readability, we will therefore use a slight abuse of notation in which we do not distinguish between the two. Decomposing $\Tgen$ and $\HConSp$ into $\SU(8)$ representations, we have 
\begin{equation}
\begin{aligned}
   \Tgen &= \rep{28} + \rep{36} + \rep{420} + \CC , \\
   \HConSp &= \left(\rep{28}+\bar{\rep{28}}\right)\times \rep{63}
        = \rep{28} + \rep{36} + \rep{420} + \rep{1280} + \CC ,
\end{aligned}
\end{equation}
and so indeed
\begin{equation}
\label{eq:SU8int}
\begin{aligned}
   \Tgen = \HTConSp &= \rep{28} + \rep{36} + \rep{420} + \CC , \\ 
   \Tint &= 0 , \\
   \HUConSp &= \rep{1280} + \CC ,
\end{aligned}   
\end{equation}
implying every $\SU(8)/\bbZ_2$ structure is torsion-free, and the space of torsion-free, compatible connections is given by $\rep{1280}+\CC$. 

Note that, much as the manifolds with special holonomy associated to covariantly constant spinors are necessarily Ricci-flat, it is easy to check that corresponding manifolds with generalised special holonomy are generalised Ricci-flat, given the definition of the generalised Ricci curvature of a generalised connection in~\cite{CSW1,CSW2,CSW3} (for the $O(d,d)\times\bbR^+$ theory these objects were first defined by Siegel~\cite{siegel}). As an example, we provide in appendix~\ref{app:ricci} a proof for the $E_{7(7)}\times\bbR^+$ case.

Our goal is now to apply this language to supersymmetric backgrounds and show that these are precisely those that satisfy the generalised special holonomy conditions.

\subsection{Simple example: the triple M5 intersection}
\label{sec:3M5}

Let us begin by considering a concrete example, the triple intersection of orthogonal M5 branes~\cite{Gauntlett:1996pb} . This solution is one of the simplest known warped four-dimensional Minkowski backgrounds preserving $\mathcal{N}=1$ supersymmetry. It therefore serves as a nice way to illustrate how generalised geometry takes supersymmetric flux solutions and repackages them as integrability conditions on the generalised tangent space. 

\subsubsection{The supergravity solution}

The metric for this supergravity solution can be written as
\begin{equation}
\label{eq:3M5-metric}
\begin{aligned}
	\dd s^2 &= (H_1 H_2 H_3)^{2/3} \Big[ (H_1 H_2 H_3)^{-1} 
            \dd s^2(\bbR^{3,1}) \\
		& \qquad 
                + H_1^{-1} \delta_{u^1 v^1} \dd x^{u^1} \dd x^{v^1} 
		+ H_2^{-1} \delta_{u^2 v^2} \dd x^{u^2} \dd x^{v^2} 
		+ H_3^{-1} \delta_{u^3 v^3} \dd x^{u^3} \dd x^{v^3} 
			+ \dd z^2 \Big] ,\\
\end{aligned}
\end{equation}
where the three harmonic functions $H_i$ depend only on the overall transverse coordinate $z$, and we define indices with the ranges
\begin{equation}
	u^1, v^1, \dots = 4,5
	\hspace{50pt}
	u^2, v^2, \dots = 6,7
	\hspace{50pt}
	u^3, v^3, \dots = 8,9
\end{equation}
Let us also define
\begin{equation}
	f_i = \log H_i \hspace{20pt} \text{ and } \hspace{20pt} \der f_i = \frac{\der f_i}{\der z} .
\end{equation}
The overall warp factor of $\dd s^2(\bbR^{3,1})$ is 
\begin{equation}
	\Delta = -\tfrac16 (f_1 + f_2 + f_3) ,
\end{equation}
and the flux is given by
\begin{equation}
\label{eq:3M5-flux}
\begin{aligned}
	F = -\ee^{2\Delta} \Big[ (\der f_1) e^{6789} + (\der f_2) e^{4589} + (\der f_3) e^{4567} \Big] ,
\end{aligned}
\end{equation}
where as usual e.g. $e^{4567} = e^4 \wedge \dots \wedge e^7$, is the wedge product of elements of the vielbein for~\eqref{eq:3M5-metric}. We see that this solution is precisely of the type described in section~\ref{sec:susyback}.

The internal components of the spin connection for the metric~\eqref{eq:3M5-metric} are
\begin{equation}
\begin{aligned}
	\tfrac14 \omega_{u^i ab} \Gamma^{ab} 
		&=-  \der_z (\Delta + \tfrac14 f_i) \Gamma_{u^i}{}^{z} ,\\
	\tfrac14 \omega_{zab} \Gamma^{ab} &= 0 ,
\end{aligned}
\end{equation}
where here the gamma matrices are still generators of the full $\Cliff(10,1)$, as for convenience we will not decompose eleven-dimensional spinors into products of internal and external spinors in this section.

%%%%%%%%%%%%%%%%%%%%%%%%%%%%%%%%%%%%

With these conventions, we have that eleven-dimensional spinors of the form
\begin{equation}
\label{eq:M5-Kspinors}
	\varepsilon = \ee^{\Delta/2} \varepsilon_0 ,
\end{equation}
with $\varepsilon_0$ a constant spinor satisfying
\begin{equation}
\label{eq:3M5-proj}
	\Gamma^{0 \dots 3} \Gamma^{45} \varepsilon_0 
		= \Gamma^{0 \dots 3} \Gamma^{67} \varepsilon_0
		= \Gamma^{0 \dots 3} \Gamma^{89} \varepsilon_0
		= \varepsilon_0 ,
\end{equation}
solve the Killing spinor equations~\eqref{eq:susy-ferm} with $d=7$. 
Note that the relations~\eqref{eq:3M5-proj} can be repackaged in the more useful form
\begin{equation}
	\Gamma^{45} \varepsilon_0
		= \Gamma^{67} \varepsilon_0 
		= \Gamma^{67} \varepsilon_0
	\hspace{30pt} \text{and} \hspace{30pt}
	\Gamma^{10} \varepsilon_0 = \varepsilon_0 .
\end{equation}
%

%%%%%%%%%%%%%%%%%%%%%%%%%%%%%%%%%%%%

\subsubsection{Generalised geometry picture}

Our claim is then that there exists a torsion-free generalised connection which is compatible with this Killing spinor.

We now want to think of spinors as being $SU(8)$ representations (or rather, $\Spin(3,1)\times\SU(8)$ representations, since in this subsection we are staying in the eleven-dimensional Clifford algebra). The natural embedding of the fermion fields involves a rescaling by the warp factor~\cite{CSW3}, namely
\begin{equation}
   \hat{\zeta} = \ee^{-\Delta/2} \zeta_{\text{sugra}},
\end{equation}
for some spinor $\zeta$. Therefore, the action of a generalised connection on the Killing spinor~\eqref{eq:M5-Kspinors} is given by
\begin{equation}
	\Dgen \hat{\varepsilon}
		=  \Dgen (\ee^{-\Delta/2} \varepsilon_{\text{sugra}}) = \Dgen \varepsilon_0 .
\end{equation}

We can use~\cite{CSW2} to read off the general form of a generalised connection $D$ compatible with $\SU(8)$ and with vanishing torsion. To compare expressions in generalised geometry with the usual supergravity ones it is convenient to express them in what is known as a conformal split frame, which provides the isomorphism $E\simeq TM\oplus\Lambda^2T^*M\oplus\Lambda^5T^*M \oplus (T^*M\otimes\Lambda^7T^*M)$, though of course the analysis can be carried out in any other $SU(8)$ frame since the language that was developed in~\cite{CSW2} is manifestly $SU(8)$ covariant. In such a frame, and given that the seven-form flux vanishes, we are left with
\begin{subequations}
\label{eq:Dgen-Q}
\begin{align}
\label{eq:Dgen-1}
	\Dgen_m \varepsilon_0 &= \ee^\Delta (\LC_m
		-\tfrac12 \tfrac{1}{4!} F_{mnpq} \Gamma^{npq} + \slashed{Q}_m) \varepsilon_0 ,\\
\label{eq:Dgen-2}
	\Dgen^{mn} \varepsilon_0 &= \ee^\Delta (\tfrac14 \tfrac{2!}{4!} F^{mn}{}_{pq} \Gamma^{pq} 
		- \tfrac1{10} (\der_p \Delta) \Gamma^{mnp} + \slashed{Q}^{mn}) \varepsilon_0 ,\\
\label{eq:Dgen-3}
	\Dgen^{m_1 \dots m_5} \varepsilon_0 
		&= \ee^\Delta (\slashed{Q}^{m_1 \dots m_5}) \varepsilon_0 ,\\
\label{eq:Dgen-4}
	\Dgen^{m,m_1 \dots m_7} \varepsilon_0 
		&= \ee^\Delta (\slashed{Q}^{m,m_1 \dots m_7}) \varepsilon_0 .
\end{align}
\end{subequations}
Here $\am \in \Gs{\HUConSp}$ are components of the connection which are unconstrained by torsion and metric compatibility. Therefore we must show that it is possible to find $\slashed{Q}$ such that $\Dgen \varepsilon_0=0$. We start simply by taking
\begin{equation}
\begin{aligned}
	\slashed{Q}^{m_1 \dots m_5} 
		&= 0 ,\\
	\slashed{Q}^{m,m_1 \dots m_7}
		&= 0 .
\end{aligned}
\end{equation}
For the remaining components, we have that they can be decomposed into irreps of $SO(7)$, of which we will only need
\begin{equation}
\label{eq:3M5-Q}
\begin{aligned}
	\slashed{Q}_m &= \tfrac14 Q_{pqm} \Gamma^{pq} ,\\
	\slashed{Q}^{mn} &= Q^{mn}{}_{pq} \Gamma^{pq}
		+ \tfrac18  Q_{pq}{}^{[m} \Gamma^{n]pq} ,
\end{aligned}
\end{equation}
with $Q_{pqm}$ and $Q^{mn}{}_{pq}$ transforming in the $\rep{105}$ and $\rep{168}$ representations of $\SO(7)$ respectively, i.e. they are completely traceless and satisfy
\begin{equation}
\begin{aligned}
	Q_{(mn)p}=Q_{[mnp]}=0 ,
	\hspace{50pt}
	Q_{mn(pq)}=Q_{(mn)pq}=Q_{[mnp]q}=0 .
\end{aligned}
\end{equation}
The coefficients of $Q_{pqm}$ in~\eqref{eq:3M5-Q} are fixed by the requirement that $Q$ drops out of the torsion of $D$. A short calculation (see appendix~\ref{app:3M5}) then shows that taking
\begin{equation}
\label{eq:Qmn}
\begin{aligned}
	Q_{u^i z v^i}  &= - 2 X_i g_{u^i v^i}   ,\\
	Q^{u^i v^j}{}_{u'^i v'^j} 
		&= \ee^{2\Delta} Y_{ij} \delta^{u^i v^j}_{u'^i v'^j}  ,
\end{aligned}
\end{equation}
with all remaining components vanishing, and where we have defined
\begin{equation}
\begin{aligned}
	Y_{ij} &= - ( A_{ij} + \tfrac14 (X_i + X_j) ) ,\\
	X_i &= - \tfrac14 \der \Delta - \tfrac18 \der f_i ,\\
	A_{ij} &= \tfrac3{20} \der \Delta + \tfrac1{24}( \der f_i + \der f_j ) - \tfrac1{24} \delta_{ij} (\der f_i) ,
\end{aligned}
\end{equation}
results in a torsion-free connection such that precisely $D\varepsilon_0=0$ like we wanted.

We have thus explicitly shown by construction that the supersymmetric triple intersection of M5 branes is an example of a manifold with generalised special holonomy. As will be explained in the next section, the reduced structure group is $\SU(7) \subset \SU(8)$, corresponding to the stabiliser of the Killing spinor $\varepsilon$.

%%%%%%%%%%%%%%

\section{General $\mathcal{N}=1$, $D=4$ supersymmetric backgrounds}
\label{sec:genbackgrounds}

Let us now return to the general case of compactifications giving  $\mathcal{N}=1$ in four dimensions. The analysis of the minimally supersymmetric backgrounds in higher dimensions is essentially identical, and is summarised in the following section. 

A key element of $\Edd\times\bbR^+$ generalised geometry is that it allows us to reformulate the Killing spinor equations in a simple $\dHd$ covariant form~\cite{CSW2,CSW3}. Focussing on $d=7$, recall that the existence of a generalised metric on a spin manifold reduces the generalised structure group to $\SU(8)$. The supergravity fermionic degrees of freedom then form $\SU(8)$ representations~\cite{deWN}. The rescaled Killing spinor parameter $\hat{\epsilon}=\ee^{-\Delta/2}\epsilon_{\text{sugra}}$ can be viewed as a section of a spinor bundle $S$, which transforms in the $\rep{8}$ representation. The rescaled internal gravitino $\hat{\psi}_m=\ee^{\Delta/2}\psi^{\text{sugra}}_m$ can be viewed as a section of a bundle $J$, which transforms in the $\rep{56}$ representation. 
The Killing spinor equations~\eqref{eq:susy-ferm} can be encoded concisely in the $\SU(8)$-covariant equations
\begin{equation}
\label{gen-susy-background}
   \delta\hat{\psi} = \Dgen \proj{J} \hat{\epsilon} = 0, \qquad 
   \delta\hat{\rho} = \Dgen \proj{S} \hat{\epsilon} = 0 .
\end{equation}
where $\Dgen$ is a generalised $\SU(8)$ connection with vanishing generalised torsion and $\proj{X}$ serves as shorthand notation for projection to the $X$ sub-bundle\footnote{Explicitly in $\SU(8)$ indices these projections are given by
\begin{equation}
    \delta\psi^{\alpha\beta\gamma} 
         = \Dgen^{[\alpha\beta} \epsilon^{\gamma]} = 0, \qquad 
    \delta \rho_{\alpha} 
         = -\bar\Dgen_{\alpha\beta} \epsilon^{\beta} = 0 .
\end{equation}
Note that in~\cite{CSW3} these projections were denoted by the symbols $\Dgen \curlywedge \hat{\epsilon}$ and $\slashed{\Dgen} \hat{\epsilon}$ respectively.}. Note that the spinor $\rho$ is related to the trace of the external gravitino. These projections define unique operators, independent of the choice of torsion-free $\SU(8)$ connection $\Dgen$. 

For both M theory~\eqref{eq:norms} and Type II backgrounds~\eqref{eq:normsII}, the norm of the Killing spinor $\hat{\epsilon}$ is constant. Hence, by definition, it is a globally non-vanishing section of $S$. The stabiliser group in $\SU(8)$ of a single element of $\rep{8}$ is simply $\SU(7)$, hence we see that, given a single Killing spinor $\hat{\epsilon}$, we have~\cite{PW}
\begin{equation*}
   \text{\emph{$\hat{\epsilon}$ defines a (global) generalised $\SU(7)$ structure}} .
\end{equation*}
In~\cite{PW} it was shown that there is an $\SU(7)$-invariant tensor constructed from spinor bilinears that is an element of the $\rep{912}$ representation and which gives an alternative definition of the structure. Here, however, we will focus only on the description in terms of spinors. The $\SU(7)$ structure unifies the different possibilities that appear when $\hat\epsilon$ was viewed as a pair of $\Spin(7)$ or $\Spin(6)$ spinors, as we now describe. 

In the M theory case, $\hat\epsilon$ defined a pair of real $\Spin(7)$ spinors, which could become parallel, meaning that at some points on the manifold the structure was $\SU(3)$ while at others it become $G_2$. In addition, one needed to keep track of the norms of the two spinor components. However, as an $\SU(8)$ object, the complex spinor always simply defines an $\SU(7)$ structure, irrespective of whether the underlying $\Spin(7)$-spinor structure is $\SU(3)$ or $G_2$. Thus in generalised geometry all supersymmetric flux backgrounds define global $G$-structures. 

The same happens for the Type II theory. There $\hat\epsilon$ defined a pair of chiral $\Spin(6)\simeq\SU(4)$ spinors, each defining an $\SU(3)$ structure. Generically they together defined an $\SU(2)$ structure, which could enhance to $\SU(3)$ if they became parallel or one vanished. Extending to $O(6,6)$ generalised geometry~\cite{GMPT}, the two spinors could be viewed as defining an $\SU(3)\times\SU(3)\subset\SU(4)\times\SU(4)\subset\SO(6,6)$ structure, which captures both the generic $\SU(2)$ case and the points where the spinors became parallel. However, points where one spinor vanished, which are stabilised by $\SU(3)\times\SU(4)\subset \SU(4)\times\SU(4)$, and describe, for example, wrapped NS fivebranes~\cite{Strominger:1986uh,chris-het,Gauntlett:2001ur}, are outside this class. Nonetheless, in all cases the combination $\hat{\epsilon}$ given in~\eqref{eq:IIspinor} defines an $\SU(7)$ structure in $\SU(8)$. The relation between the stabilised groups in the $E_{7(7)}\times\bbR^+$ and $O(6,6)\times\bbR^+$ generalised geometries can be viewed as follows  
\begin{equation*}
   \begin{array}{ccccccccc}
      E_{7(7)} && \supset && \SU(8) && \supset && \SU(7) \\
         \cup &&&& \cup &&&& \cup \\
      \Spin(6,6) && \supset &&  \SU(4)\times\SU(4) 
         && \supset && \SU(3)\times\SU(3) 
   \end{array} . 
\end{equation*}
The $\SO(6,6)$ pure spinors $\Phi^\pm$ embed as $E_{7(7)}\times\bbR^+$ tensors transforming in the $\rep{56}$ and $\rep{133}$ representations~\cite{GLSW}. Neither is $\SU(7)$ invariant and so they are not generically globally defined. However taking the $\rep{912}$ representation in their product $\rep{56}\times\rep{133}$ gives the $\SU(7)$ invariant of~\cite{PW}. 

The condition for a generalised connection to be compatible with this $\SU(7)$ structure is then\footnote{Note that if the norm of $\hat{\epsilon}$ was not constant but was nowhere vanishing, it would still define an $\SU(7)$ structure, but the condition would become $\hat{\Dgen}\hat{\eta}=0$ where $\hat{\eta}=\hat{\epsilon}/\sqrt{\bar{\hat{\epsilon}}\hat{\epsilon}}$ is the rescaled unit norm object.} $\hat{\Dgen}\hat{\epsilon}=0$. The equations~\eqref{gen-susy-background}, which hold for any torsion-free $\SU(8)$ compatible $\Dgen$, appear weaker than the compatibility condition, as they constrain only two of the irreducible $\SU(8)$ projections of $\Dgen \hat\epsilon$. However, we will show that if~\eqref{gen-susy-background} holds one can still find a torsion-free connection $\hat\Dgen$ such that $\hat\Dgen \hat\epsilon= 0$.  As we saw in section~\ref{sec:TintGG}, this means that the generalised intrinsic torsion of the $SU(7)$ structure vanishes. Conversely, if $\hat\Dgen \hat\epsilon= 0$ then the Killing spinor conditions~\eqref{gen-susy-background} are satisfied trivially, and so we have that
\begin{quote}
   \textit{The Killing spinor equations are equivalent to the vanishing of generalised intrinsic torsion. Supersymmetric four-dimensional $\mathcal{N}=1$ backgrounds are in one-to-one correspondence with manifolds of $SU(7)$ generalised special holonomy.} 
\end{quote}
The proof turns out to be rather simple, involving only basic linear algebra.

%%%%%%%%%%%%%%%%%%%%%%%%%%%%%%%%%%

\subsection{Generalised torsion-free $\SU(7)$ structures}
\label{sec:proof}

In the following, as in section~\ref{sec:TintGG}, for the sake of readability we will use a slight abuse of notation in which we identify bundles with their corresponding representations. 

Let us start by calculating the generalised intrinsic torsion of a $\SU(7)$ structure. Decomposing into $\SU(8)\supset\SU(7)$ representations we have the torsion
\begin{equation}
\begin{aligned}
   \Tgen &= \rep{28} + \rep{36} + \rep{420} + \CC , \\ 
       &= (\rep{7} + \rep{21}) 
           + (\rep{1} + \rep{7} + \rep{28}) 
           + (\rep{21} + \rep{35} + \rep{140} + \rep{224} ) 
           + \CC ,
\end{aligned}
\end{equation}
while for $\Kgen_{\SU(7)}$ we have 
\begin{equation}
\begin{aligned}
   \Kgen_{\SU(7)} &= (\rep{7} + \rep{21} + \CC) \times \rep{48} , \\
       &= \rep{7} + \rep{21} + \rep{28} 
           + \rep{140} + \rep{189} + \rep{224} + \rep{735}
           + \CC . 
\end{aligned}
\end{equation}
Thus it appears that the space of intrinsic torsions is given by those representation in $\Tgen$ that do not appear in $\Kgen_{\SU(7)}$, namely  
\begin{equation}
\label{eq:SU7Tint}
   \Tint = \rep{1} + \rep{7} + \rep{21} + \rep{35} + \CC . 
\end{equation}

However, it could be that the kernel of the map $\tau:\Kgen_{\SU(7)}\to W$ is more than just the $\rep{189}$ and $\rep{735}$ representations, and hence $\Tint$ is actually larger. To see that this is not the case we need the explicit map. In $\SU(8)$ indices, sections $\hat\Sigma\in\Gs{\HConSp}$ are given by
\begin{equation}
   \hat\Sigma = (\hat\Sigma_{\alpha\beta}{}^\gamma{}_\delta , 
        \bar{\hat\Sigma}^{\alpha\beta}{}^\gamma{}_\delta , )
        \in (\rep{28}+\bar{\rep{28}})\times\rep{63} = \Kgen_{SU(8)}, 
\end{equation}
where the elements are antisymmetric on $\alpha$ and $\beta$ and traceless on contracting $\gamma$ with $\delta$. The map $\tau$ is then given, up to overall normalisations, by 
\begin{equation}
\begin{aligned}
   \tau(\hat\Sigma)_{\alpha\beta} 
      &= \hat\Sigma_{\alpha\gamma}{}^\gamma{}_{\beta} , 
      && \in \rep{36} + \rep{28} , \\
   \tau(\hat\Sigma)_{\alpha\beta\gamma}{}^\delta 
      &= \hat\Sigma^0_{[\alpha\beta}{}^\delta{}_{\gamma]} , 
      && \in \rep{420} , 
\end{aligned}
\end{equation}
where the ``0'' superscript on $\hat\Sigma^0_{[\alpha\beta}{}^\delta{}_{\gamma]}$ means it is completely traceless. The $\rep{28}$ and $\rep{36}$ representation just correspond to the symmetric and anti-symmetric parts of $\tau(\hat\Sigma)_{\alpha\beta}$. There are similar expressions for the conjugate representations in terms of $\bar{\hat\Sigma}$. 

If we now turn to $\SU(7)$ compatible connections and write $\Sigma\in \Gs{\Kgen_{SU(7)}}$, we can split the spinor indices $\alpha$ into $a=1,\dots,7$ and $8$ so that the non-zero components are
\begin{equation}
\begin{aligned}
   \Sigma_{ab}{}^c{}_d &\in \rep{21}\times\rep{48} , \\
   \Sigma_{a8}{}^c{}_d = -\Sigma_{8a}{}^c{}_d
       &\in \rep{7}\times\rep{48} , 
\end{aligned}
\end{equation}
and similarly for the conjugate $\bar{\Sigma}$. We then find
\begin{equation}
\label{eq:map1}
\begin{aligned}
   \tau(\Sigma)_{ab} &= \Sigma_{ac}{}^c{}_b  && \in \rep{21}+\rep{28} , & 
      && && && 
   \tau(\Sigma)_{a8} &= 0  &&\in \rep{7} , \\
   \tau(\Sigma)_{8b} &= \Sigma_{8c}{}^c{}_b  &&\in \rep{7} , & 
      && && && 
   \tau(\Sigma)_{88} &= 0 &&\in \rep{1} , 
\end{aligned}
\end{equation}
and 
\begin{equation}
\label{eq:map2}
\begin{aligned}
   \tau(\Sigma)_{abc}{}^d &= \Sigma^0_{[ab}{}^d{}_{c]}  
      && \in \rep{224} , & 
      && && && 
   \tau(\Sigma)_{abc}{}^8 &= 0  &&\in \rep{35}, \\
   \tau(\Sigma)_{ab8}{}^c &= \Sigma^0_{8[a}{}^c{}_{b]}  
      &&\in \rep{140} ,& 
      && && && 
   \tau(\Sigma)_{ab8}{}^8 &= 0  &&\in \rep{21} . 
\end{aligned}
\end{equation}
This verifies explicitly that $\Tint$ is indeed given by~\eqref{eq:SU7Tint}.

\subsection{Killing spinors and generalised intrinsic torsion}
\label{sec:KSgenint}

We now turn to showing that the Killing spinor equations set the intrinsic torsion of the $\SU(7)$ structure to zero. Decomposing into $\SU(7)$ representations we have
\begin{equation}
   \Dgen \proj{J} \hat{\epsilon} \in \rep{35} + \rep{21} , 
   \qquad 
   \Dgen \proj{S} \hat{\epsilon} \in \rep{7} + \rep{1} , 
\end{equation}
and hence the Killing spinor equations transform in the same complex representations that appear in $\Tint$, so it is reasonable that they imply that the structure is torsion-free. 

To see in detail that this is indeed the case, we first note that the compatible $\SU(7)$ connection $\hat{\Dgen}$ must also be an $\SU(8)$ connection and hence can be written as 
\begin{equation}
   \hat{\Dgen} = \Dgen + \hat{\Sigma} ,
\end{equation}
where $\hat{\Sigma}\in\HConSp$. Since $\Dgen$ is torsion-free we have
\begin{equation}
\label{eq:TS}
   T(\hat{D}) = \tau(\hat{\Sigma}) . 
\end{equation}
By definition $\hat{\Dgen}\hat{\epsilon}=0$ so in particular the projections $\hat{\Dgen}\proj{J}\hat{\epsilon}$ and $\hat{\Dgen}\proj{S}\hat{\epsilon}$ both vanish. Thus we have 
\begin{equation}
   \hat{\Sigma} \proj{J} \hat{\epsilon} = 0 , \qquad 
   \hat{\Sigma} \proj{S} \hat{\epsilon} = 0 .
\end{equation}
In spinor indices these read
\begin{equation}
\begin{aligned}
   \hat{\Sigma}_{[ab}{}^8{}_{c]} = \tau(\hat{\Sigma})_{abc}{}^8 &= 0 
   & & \in \rep{35} , & && && 
   \hat{\Sigma}_{[8a}{}^8{}_{b]} = \tau(\hat{\Sigma})_{ab8}{}^8 & = 0  
   & & \in \rep{21} , \\
   \hat{\Sigma}_{ab}{}^b{}_8 = \tau(\hat{\Sigma})_{a8} &= 0 
   & & \in \rep{7} , & && && 
   \hat{\Sigma}_{8a}{}^a{}_8 = \tau(\hat{\Sigma})_{88} & = 0  
   & & \in \rep{1} , 
\end{aligned}
\end{equation}
together with their complex conjugates. But, given~\eqref{eq:TS} and  comparing with~\eqref{eq:map1} and~\eqref{eq:map2}, we see that this precisely sets the intrinsic torsion of $\hat{D}$ to zero. 

Since, by definition, if $\hat{D}$ is a torsion-free $\SU(7)$ connection then~\eqref{gen-susy-background} are satisfied, we have shown that the Killing spinor equations are equivalent to the existence of a torsion-free $\SU(7)$ structure, and hence imply the manifold has $\SU(7)$ generalised special holonomy.  

Note that, given~\eqref{eq:SU8int}, the space of $\SU(8)$ connections decomposes as 
\begin{equation}
   \HConSp = \Tgen \oplus \HUConSp 
      = \Tint \oplus \Tgen_{\SU(7)} \oplus \HUConSp . 
\end{equation}
The calculation above shows that the map
\begin{equation}
\begin{aligned}
   \mathcal{P}^{\epsilon} :\HConSp &\rightarrow 
        (S \oplus J) \oplus \overline{(S \oplus J)},\\
   \hat{\Sigma} &\mapsto (\hat{\Sigma} \proj{S} \epsilon) 
      + (\hat{\Sigma} \proj{J} \epsilon) + \CC ,
\end{aligned}
\end{equation}
when restricted to $\Tint$, defines an isomorphism 
\begin{equation}
   \Tint \simeq (S \oplus J) \oplus \overline{(S \oplus J)} .
\end{equation}
As we will see in the next section, this same structure will appear for minimally supersymmetric compactifications to five- and higher dimensional Minkowski space. 

\section{On minimally supersymmetric backgrounds in $D>4$}
\label{sec:genD}

So far we have narrowed our discussion to four-dimensional Minkowski backgrounds, corresponding to internal seven-dimensional manifolds (or six-dimensional in the Type II case) which are described by $E_{7(7)}\times\bbR^+$ generalised geometry. However, using the results of~\cite{CSW2,CSW3} the analysis of sections~\ref{sec:proof} and~\ref{sec:KSgenint} can be used, mutatis mutandis, to show that, for all compactifications to $D\geq4$, preserving minimal supersymmetry is equivalent to the internal manifold having generalised special holonomy. 

In the following we will not give the details of the analysis, since they are straightforward, but simply summarise the groups and representations that appear. As discussed in section~\ref{sec:gg}, the formalism developed in~\cite{CSW2,CSW3} reformulated supergravity restricted to manifolds with $d\leq7$ dimensions for M theory (or $d-1\leq6$ dimensions in the Type II case), using the corresponding $\Edd\times\bbR^+$ generalised geometry. In each case, the (rescaled) Killing spinor $\hat{\epsilon}$ has fixed norm, and so defines a $G\subset\dHd$ structure. These are listed in table~\ref{table}. One can again calculate the $G$ representations that appear in the intrinsic torsion and compare these to the representation of spinor bundles $S$ and $J$. In the table~\ref{table} the latter are given as $\dHd$ representations. 
\begin{table}[htb]
\centering
\begin{tabular}{llllll}
   $d$ & $\dHd$ & $S$ & $J$ & $G$ & $\Tint$ \\[3pt]
   \hline \\[-12pt]
   7 & $\SU(8)$ & $\rep{8}$ & $\rep{56}$ 
      & $\SU(7)$ & $\rep{1} + \rep{7} + \rep{21}+ \rep{35} + \CC$\\[3pt]
   6 & $ \Symp(8)$ & $\rep{8}$ & $\rep{48}$ & $\Symp(6)$ 
      & $2\cdot\rep{1} + 2\cdot\rep{6} + 2\cdot\rep{14}$ \\
      &&&&&\quad ${}+ \rep{14'} + \CC $\\[3pt]
   5 & $\Symp(4)^2$ & $\repp{4}{1} + \repp{1}{4}$ 
      & $\repp{4}{5} + \repp{5}{4}$ & $\Symp(2)\cdot\Symp(4)$
      & $2\cdot\repp{1}{4} + 2\cdot\repp{2}{4} + \CC $ \\[3pt]
   4 & $\Symp(4)$ & $\rep{4}$ & $\rep{16}$	
      & $\Symp(2)$ 
      & $4\cdot\rep{1} + 5\cdot\rep{2} + 2\cdot\rep{3} + \CC$
\end{tabular}
\caption{Representations of the bundles $S$ and $J$, and the space of intrinsic torsions of the generalised $G$-structure defined by a globally non-vanishing section of $S$. Note that $\Symp(2n)$ denotes the compact symplectic group of rank $n$.}
\label{table}
\end{table}

In each case, the $S$ and $J$ representations are complex. As in section~\ref{sec:KSgenint} we see that there is an isomorphism 
\begin{equation}
   \Tint \simeq (S \oplus J) \oplus \overline{(S \oplus J)} ,
\end{equation}
so that imposing the Killing spinor equations precisely sets the generalised intrinsic torsion to zero. There is a subtlety that arises for $d=5$, since in that case the spinor bundles $S$ and $J$ are reducible. We write $S=S^+ \oplus S^-$ where the fibre of $S^+$ is the $(\rep{4},\rep{1})$ representation while the fibre of $S^-$ is the $(\rep{1},\rep{4})$ representation. Similarly we have the split $J=J^+\oplus J^-$ where $J^+$ and $J^-$ correspond to the $(\rep{4},\rep{5})$ and $(\rep{5},\rep{4})$ parts respectively.
If a spinor is a section of $S^+$ then the corresponding fermion supersymmetry variations transform in $S^-\oplus J^-$. If the minimal $\mathcal{N}=(1,0)$ supersymmetry is preserved,  
the Killing spinor derivatives are sections of $S^-\oplus J^-$ and we see that the decomposition of this, with its complex conjugate, again matches $\Tint$. Note also that the $d=4$ case is simply that $M$ has conventional $\SU(2)\simeq\USp(2)$ holonomy, since the flux $F$ and warp factor $\Delta$ are necessarily zero in this case. The same is true for $d\leq 3$. 

We arrive at the final result:
\begin{quote}
   \textit{The minimally supersymmetric Minkowski backgrounds with $D\geq 4$ are in one-to-one correspondence with manifolds with generalised     special holonomy group $G$ where $G=\SU(7), \USp(6), \USp(2)\times\USp(4), \USp(2)$ in dimensions $D=4,5,6,7$ respectively.}
\end{quote}
%

%%%%%%%%%%%%%%%%%%%%%%%%%%%%%%%%%%%%%%%%%%%%%%%%%%%%%%%%%%%%%%%%%%%%%%%%%%

\section{Discussion}
\label{sec:conc}

We have found a novel integrability condition that describes all the
internal manifolds resulting in minimally supersymmetric
compactifications of M theory and Type II theory to $D$-dimensional Minkowski space for $D\geq 4$ -- they are manifolds with a generalised $G$-structure that has vanishing intrinsic torsion. One can think of these spaces as the generalised geometry analogues of special holonomy manifolds, now with general fluxes included.

This reformulation gives a new geometric understanding of such backgrounds. For example, it is easy to show that manifolds with a generalised special holonomy defined by Killing spinors are generalised Ricci-flat, as defined in~\cite{CSW1,CSW2}. This amounts to a geometric restatement of the supergravity result that the Killing spinor equations together with the Bianchi identities solve the equations of motion for the Minkowski backgrounds~\cite{Gauntlett:2002fz,Gauntlett:2002sc,Gauntlett:2003cy,
Grana:2006kf,Koerber:2007hd}. It is a good illustration of the power of a formalism with full manifest symmetries -- a supergravity result which requires rather arduous computations has been simplified to an argument expressed in a couple of lines. In particular, the fact that none of the flux equations of motion have to be separately imposed is conventionally either a lengthy derivation~\cite{Koerber:2007hd} or seen only given a particular ansatz for the structure~\cite{Gauntlett:2002fz,Gauntlett:2003cy}. 

This classification is also complete. Solutions like, for example for $d=7$, the wrapped NS fivebrane manifold~\cite{Strominger:1986uh,chris-het,Gauntlett:2001ur} which fall outside the formulation of~\cite{GMPT}, are contained in the classification here. It also gives a new geometric interpretation of the results of~\cite{GO,GO2}. These papers had already shown that the Killing spinor equations could be written as differential conditions in the language of generalised geometry. They considered the $E_{7(7)}$ tensors, introduced in~\cite{GLSW}, that are the lifts of the pure spinors $\Phi^\pm$ to exceptional generalised geometry, and that define $E_{6(2)}\subset E_{7(7)}$ and $\SO^*(12)\times\SU(2)\subset E_{7(7)}$ structures. They then showed that these objects satisfy differential constraints given by twisted operators transforming in the $\rep{56}+\rep{912}$ representations of $E_{7(7)}$. These are none other than the generalised torsion representations -- in fact, we now recognise that these differential conditions amount precisely to setting the intrinsic torsion of the generalised $SU(7)$ structure to zero.

It is also interesting to contrast the notion of generalised special holonomy given here with that in~\cite{Duff:1990xz,Berkooz:1996km,FigueroaO'Farrill:2002ft,gen-holo-duff,Hull:2003mf,Lu:2005im}. In these earlier works, one considered the conventional holonomy of the non-generic Clifford connection that appears in the gravitino variation in~\eqref{eq:susy-ferm}. Here we are considering a generalised connection $\Dgen$. The object that appears in the gravitino variation is not the connection itself but a projection~\eqref{gen-susy-background}. Furthermore, this operator is uniquely defined by its geometrical properties, namely that it is torsion-free and compatible. It precisely because we consider generalised derivatives that we can formulate the notion of integrability which determines the supersymmetric backgrounds.

There are a number of ways in which these results can be extended. First, one can consider any number of torsion-free generalised structures, not only those defined by Killing spinors. For example, in the original $O(d,d)\times\bbR^+$ generalised geometry, it is easy to see that for $d=2n$, generalised complex structures and generalised Calabi--Yau structures are indeed torsion-free $U(n,n)$ and $\SU(n,n)$ structures respectively. One can similarly define torsion-free conditions for the $E_{6(2)}$ and $\SU(2)\times\SO^*(12)$ structures introduced in~\cite{GLSW}. These define the full M theory or Type II generalisations of conventional complex and symplectic structures in six dimensions. It is also straightforward to adapt our discussion to case where $M$ is a Lorentzian manifold and the warped factor in~\eqref{eq:g} is Euclidean. The generalised structure groups remain $\Edd\times\bbR^+$ but the Lorentzian generalised metric defines a subgroup $\HdL$ with a different signature from $\Hd$ that appears in the Euclidean case~\cite{Hull:1998br,gen-holo-duff,Hull:2003mf}, so that, for example $\SU(8)$ becomes $\SU^*(8)$. Supersymmetry will again correspond to special holonomy with $G\subset\HdL$. As in the case of conventional Lorentzian special holonomy~\cite{FigueroaO'Farrill:1999tx,Bryant,Leistner}, since $\HdL$ is non-compact the stabliser groups of the Killing spinor can be more complicated, and in addition, it is no longer guaranteed that the equations of motion are satisfied. 

Not only can one consider generalised complex and Calabi--Yau structures in the $O(d,d)\times\bbR^+$ case, but, using~\cite{CSW1}, the intrinsic torsion formalism can also be applied directly in ten dimensions with Lorentzian signature.
We anticipate that a Killing spinor in ten dimensions will again define an integrable generalised $G$-structure on $TM\oplus T^*M$ for backgrounds with only NSNS fluxes. In particular, an extension of the pure spinor approach to supersymmetric vacua in generalised complex geometry to ten dimensions has appeared in~\cite{Tomasiello:2011eb}, and been used to analyse Lifshitz solutions in~\cite{Petrini:2012bh}. (Generalised complex geometry has also been applied to ``pure'' Euclidean ten-dimensional vacua in~\cite{Prins:2014ona}.) It would be interesting to see exactly how the conditions formulated in~\cite{Tomasiello:2011eb} correspond to vanishing generalised intrinsic torsion in our formalism. 

One can also consider structures in other versions of generalised geometry. An obvious case is the Strominger system, that is the $D=4$, $\mathcal{N}=1$ heterotic compactification with $H$-flux, first considered in~\cite{Strominger:1986uh,chris-het}. If $\dd H=0$ it can be described in Type II, and is simply an $\SU(3)\times\SU(4)\subset\SU(7)$ special holonomy manifold with one $\epsilon_i$ vanishing, namely, the wrapped NS fivebrane background discussed above. To incorporate the gauge fields and consider $\dd H\neq0$ one must extend the $O(d,d)$ generalised tangent space~\cite{DFT-het,GM,GF,Baraglia,Coimbra:2014qaa}, a construction that is closely related to $B_n$ generalised geometry discussed in the mathematics literature~\cite{Bn}. Focussing on the construction of~\cite{Coimbra:2014qaa}, the generalised metric defines a local symmetry group $H_d^{\text{het}}=O(d)\times O(d)\times G$ where $G$ is the gauge group. However, the corresponding structure is not torsion-free but has some very particular intrinsic torsion. Simply from the form of the Killing spinor equations it seems likely that the Strominger system should correspond to an $\SU(3)\times\SU(4)\times G\subset H_6^{\text{het}}$ structure, with the same intrinsic torsion as the generalised metric structure. This is the closest notion one has in this context to being special holonomy.  In another direction, in~\cite{CSC}, several classes of new generalised geometries were introduced, including ``half-exceptional'' generalised geometries such as $Spin(8,8)\times\bbR^+$ generalised geometry that can be used to describe compactifications of M theory to three dimensions with seven-form flux.\footnote{This avoids the issues~\cite{CSW2} present on attempts to formulate a hypothetical $E_{8(8)}\times\bbR^+$ generalised geometry which would contain all the internal bosonic fields in a three-dimensional background (though see also~\cite{Berman:2012vc,EFT1b,Rosabal:2014rga}).} The internal manifolds of $\mathcal{N}=1$ vacua in that setup would possess an integrable generalised $\Spin(7) \times \Spin(8)$ or $\Spin(7) \times \Spin(7)$ structure depending on whether the internal supersymmetry parameter is chiral or non-chiral respectively. (The interpolating cases are not combined as single generalised structure groups in $\Spin(8)\times\Spin(8)$). Similarly, using other generalised geometries sketched in~\cite{CSC}, one could consider the corresponding structures for compactifications of certain six-dimensional $\mathcal{N}=(1,0)$ supergravity theories.

Finally, two very obvious extensions are the description of AdS backgrounds and backgrounds with more supersymmetry. For AdS, we can show the cosmological constant will appear as a constant scalar parameter defining the generalised intrinsic torsion of the $G$-structure. For the standard examples, describing branes at conical singularities and where only a top-form flux is present, there is indeed a conventional torsionful $G$-structure~\cite{Acharya:1998db}. Decomposing under $G$ only the singlet components of the intrinsic torsion are non-zero (and constant), defining, for example, weak-$G_2$ and Sasaki--Einstein metrics. By modifying the arguments of section~\ref{sec:proof} we can easily verify that exactly the same happens in the generalised case: the singlet parts of the intrinsic torsion are non-zero. This is summarised in table~\ref{tab:AdS}. 
\begin{table}[htb]
\centering
\begin{tabular}{lllll}
	$d$ & $G$ & $G_{\text{com}}$ & R-symmetry &  $\TTint$  \\ 
	\hline 
	7 & $\SU(7)$ & $ U(1)$ &  $\bbZ_2$ & $\rep{1}_2$ \\
	6 & $ \Symp(6)$ &$ \USp(2)$ &  $U(1)$ & $\repp{3}{1}$ \\
	5 & $\Symp(2)\times\Symp(4)$ & $\USp(2)$ & ---
             & no singlets \\
	4 & $\Symp(2)$ & $\USp(2)$  &$\USp(2)$ & $2\cdot\repp{1}{1}$
\end{tabular}
\caption{Generalised structure subgroups $G\subset \dHd$, commutant groups $G_{\text{com}}$ of $G$ in $\dHd$, AdS R-symmetry groups and non-vanishing generalised intrinsic torsion as representations of $G_{\text{com}}\times G$ for minimal supersymmetry in AdS backgrounds.}
\label{tab:AdS}
\end{table}
We first note, looking at table~\ref{table}, that the minimal supersymmetric structure for $d=5$ has no singlets in $\Tint$. This would imply it does not allow an AdS background. This is the standard result that there is no AdS solution in six dimensions preserving $\mathcal{N}=(1,0)$ supersymmetry~\cite{Nahm:1977tg}. For all the other cases, $\Tint$ does contain singlets. To see which singlets correspond to the AdS torsion we consider the R-symmetry of the AdS group. This is always a subgroup of R-symmetry of the flat-space background, which is itself simply the commutant $G_{\text{com}}$ of the structure group $G$ in $\Hd$. These are both listed in the table. The AdS intrinsic torsion should break the $G_{\text{com}}$ to the AdS R-symmetry. Keeping track of their representations under $G_{\text{com}}\times G$, as given in the table, this is enough to fix the correct singlets in $\Tint$. In the Killing spinor equations on AdS the intrinsic torsion appears as the cosmological constant term in the external gravitino variation, for example, as a constant, charge-two superpotential in $D=4$, and as a triplet of prepotentials in $D=5$. The $D=7$ case is interesting in that in M theory there are no smooth $\mathcal{N}=1$ AdS backgrounds~\cite{Acharya:1998db}. Instead, the only possibility is the maximal $\mathcal{N}=2$ solution on $S^4$. The AdS R-symmetry is $\USp(4)$ in this case, and there is then only one singlet in the intrinsic torsion~\cite{Lee:2014mla}.  However, recently a family of $\mathcal{N}=1$ solutions in massive Type IIA theory was discovered~\cite{Apruzzi:2013yva}, and these should fall into the class discussed here. Recall that the cone spaces over the classical Sasaki--Einstein and weak $G_2$ spaces are special holonomy manifolds. The same will happen here. Viewing the $D$-dimensional AdS space as a supersymmetric warped compactification to $(D-1)$-dimensional flat space, implies that the cones over the spaces listed in table~\ref{tab:AdS} must all be special holonomy spaces for $E_{d+1(d+1)}\times\bbR^+$ generalised geometry. 

The second question is to ask what happens if we preserve more supersymmetry. Again the natural conjecture is that these are all generalised special holonomy manifolds. The possible holonomy groups are as given in table~\ref{table2}. 
\begin{table}[htb]
\centering
\begin{tabular}{lll}
   $d$ & $\dHd$ &  $G_{\mathcal{N}}$  \\ 
   \hline
   7 & $\SU(8)$ &  $\SU(8-\mathcal{N})$\\
   6 & $ \Symp(8)$ & $\Symp(8-2\mathcal{N})$ \\
   5 & $\Symp(4)\!\! \times \!\! \Symp(4)$ &   
      $\Symp(4-2\mathcal{N}_+)\!\times\!\Symp(4-2\mathcal{N}_-)$\\
   4 & $\Symp(4)$ & $\Symp(4-2\mathcal{N}) $
\end{tabular}
\caption{Generalised structure subgroups $G_{\mathcal{N}}\subset \dHd$ preserving $\mathcal{N}$ supersymmetry in $(11-d)$-dimensional Minkowski backgrounds. Note that for $d = 5$ we have six-dimensional supergravity with $(\mathcal{N}_+,\mathcal{N}_-)$ supersymmetry.}
\label{table2}
\end{table}
That the Killing spinors define these structures follows immediately from the same arguments as in section~\ref{sec:genbackgrounds}.  Proving that the Killing spinor equations imply they are torsion-free is generally less straightforward and cannot be given just in terms of simple representation theory as we have done in this paper. (There is also of course a key additional question of when examples of such holonomies exist.) Focussing in $d=7$, the exception is the $\mathcal{N}=2$ case, first considered in~\cite{GLSW,GO2}, for which one can simply go through the exact same steps of section~\ref{sec:proof}, except using an $\SU(6)$ decomposition together with the fact that now one has two copies of the Killing spinor constraints. One can thus think, from a string theory perspective, of manifolds with $SU(6)$ generalised special holonomy as the direct analogue of Calabi-Yau threefolds. They are in a sense the final step in extending these celebrated spaces, a kind of ``exceptional generalised Calabi-Yau'' manifold, 
\begin{equation*}
\begin{array}{llll}
   \text{\emph{Manifold}} & \text{\emph{Integrable Structure} \quad} 
   & \text{\emph{Tangent Space} \quad} &  \text{\emph{$\mathcal{N}=2$ background}}  \\*[5pt] 
   \text{Calabi-Yau} & \SU(3) & TM &  \text{Fluxless} \\
   \text{Generalised CY} & \SU(3)\times\SU(3) & TM\oplus T^*M & \text{NSNS sector} \\
   \text{Exceptional GCY \quad} & \SU(6) & E & \text{Generic}
\end{array}
\end{equation*}
More generally, for higher $\mathcal{N}$ the argument is more subtle, and requires introducing some new concepts to generalised geometry. We hope to present these results in a forthcoming paper.
 
%%%%%%%%%%%%%%%%%%%%%%%%%%%%%%%%%%%%%%%%%%%%%%%%%%%%%%%%%%%%%%%%%%%%%%%%%%

\acknowledgments

This work was supported in part by the Portuguese Funda\c c\~ao para a Ci\^encia e a Tecnologia under grant SFRH/BD/43249/2008 and the German Research Foundation (DFG) within the Cluster of Excellence ``QUEST'' (A.~C.), the German Science Foundation (DFG) under the Collaborative Research Center (SFB) 676 ``Particles, Strings and the Early Universe'' (C.~S.-C.), and the EPSRC Programme Grant ``New Geometric Structures  from String Theory'' EP/K034456/1 and the STFC Consolidated Grant ST/J0003533/1 (D.~W.). D.~W. also thanks the Berkeley Center for Theoretical Physics for kind hospitality during the final stages of this work.

%%%%%%%%%%%%%%%%%%%%%%%

\appendix

\section{$G$-structures and intrinsic torsion}
\label{app:Tint}

We briefly review the concepts of $G$-structures and intrinsic torsion, following the approach of~\cite{Joyce-Book}. A $G$-structure on a $d$-dimensional manifold is a principal sub-bundle $P_G\subset F$, with fibre $G$, of the $\GL(d;\bbR)$ frame bundle $F$ of $M$. The existence of this sub-bundle is typically equivalent to specifying a set of globally defined $G$-invariant tensors or, as most relevant here, when $G\subset \Spin(d)$, a metric and a set of globally defined $G$-invariant spinors. 

One can always find a connection $\hat{\nabla}$ that is compatible with the $G$-structure. This means that the corresponding connection on the principal bundle $F$ reduces to a connection on $P_G$. Equivalently, given a basis $\{\hat{e}_a\}$ of $TM$ in $P_G$, one has a set of connection one-forms $\omega^a{}_b$ taking values in the adjoint representation of $G$. If the structure is defined by a $G$-invariant tensor $\Phi$, compatibility is then also equivalent to having $\hat{\nabla}\Phi = 0$. 

It follows that if one takes two arbitrary connections $\hat{\nabla}$ and $\hat{\nabla}'$ which are compatible with $P_G$, their difference defines a tensor which is section of $T^*M \otimes \adj P_G$, so that 
\begin{equation}
   \Sigma  = \hat{\nabla}' - \hat{\nabla}, 
   \qquad
   \Sigma \in \Gs{ T^*M \otimes \adj P_G}.
\end{equation}
The torsion of a generic connection in turn will be a section of the bundle
\begin{equation}
   T(\hat{\nabla})\in \Gs{TM \otimes \Lambda^2 T^*M},
\end{equation} 
Typically both of these tensor product bundles can be decomposed into irreducible parts under $G$.

The intrinsic torsion of $P_G$ can then be defined as follows. Let us label these two bundles as 
\begin{equation}
   \GConSp=T^*M \otimes \adj P_G, 
   \qquad 
   \Tgen=TM \otimes \Lambda^2 T^*M.
\end{equation}
One then defines a map $\tau$ between them
\begin{equation}
   \tau: \GConSp \rightarrow \Tgen,
\end{equation} 
given by
\begin{equation}
   \tau(\Sigma) =  T(\hat{\nabla}') - T(\hat{\nabla}).
\end{equation}
It can be the case that the image of this map does not fill out the full space of torsions $\Tgen$. Denoting the vector bundle associated to the image of $\tau$ by $\im\tau=\GTConSp$, we can therefore define
\begin{equation}
	\Tint = \Tgen/\GTConSp .
\end{equation}
By construction $\Tint$ does not depend on the choice of compatible connection, only on the structure $P_G$. In other words, if we project the torsion of any compatible connection $T(\hat{\nabla})$ onto $\Tint$, we obtain an element $\TTint(P_G)\in\Gs{\Tint}$ which is independent of the connection we chose. This element of $\TTint$ is called the intrinsic torsion of $P_G$, and if it is non-zero, then there does not exist a torsion-free connection which is compatible with $P_G$.  Typically (for example if $G\subset O(d)$), we can decompose $\Tint$ 
\begin{equation}
\Tint = \bigoplus_{(i)} W_{(i)},
\end{equation}
with the fibres of the $W_{(i)}$ being associated to irreducible representations of $G$, which can be useful for classification purposes. 

If the intrinsic torsion vanishes we can find a torsion-free $\hat{\nabla}$ compatible with $G$ and then we say the $G$-structure is integrable or torsion-free. If in addition $G\subset O(d)$, since the Levi--Civita connection $\LC$ is the unique torsion-free $O(d)$ connection, we must have $\hat{\nabla}=\LC$, and hence $M$ is a manifold with special holonomy group $G$. 

In general, integrability is a first-order differential constraint on the structure. Consider the case $G\subset O(d)$ and suppose the structure is defined by a $G$-invariant tensor $\Phi$. Let $\hat{\nabla} = \LC + \hat\Sigma$, where this time the Levi--Civita connection $\LC$ is  torsion-free (but not necessarily compatible) and $\hat{\nabla}$ is a connection compatible with $\Phi$ (but not necessarily torsion-free). In other words, we have that
\begin{equation}
   0 = \hat{\nabla} \Phi = \LC \Phi + \hat\Sigma \cdot \Phi ,
\end{equation}
where $\hat\Sigma\in \Gs{T^*M\otimes\adj P_{O(d)}}$. Since $\nabla$ is the unique torsion-free $O(d)$ connection, we see that the obstruction to $\hat{\nabla}$ being torsion-free is $\hat\Sigma \cdot \Phi$. Decomposing $\adj P_{O(d)}\simeq \Lambda^2T^*M = \adj P_G \oplus \adj P_G^\perp$, since the action of $\adj P_G$ on $\Phi$ is trivial by definition, we see that in this case we can identify  
\begin{equation}
   \Tint \simeq T^*M\otimes \adj P_G^\perp , 
\end{equation}
and view $\hat\Sigma\cdot\Phi=-\LC\Phi$ as the intrinsic torsion $\TTint$. 

As an example, consider an M-theory flux background on a seven-dimensional manifold admitting a single Killing spinor defining an $G_2$-structure. (A similar discussion of the intrinsic torsion classes of the $\SU(3)$ structure discussed in section~\ref{sec:Gstruct} is given in~\cite{Dall'Agata:2003ir}.) The adjoint representation of $\SO(7)$ decomposes under $G_2$ as $\rep{21} \rightarrow \rep{14}+\rep{7}$, so a quick computation gives the intrinsic torsion space in terms of four irreducible components
\begin{equation}
   \Tint = W_1 \oplus W_7 \oplus W_{14} \oplus W_{27} ,
\end{equation}
with respective fibers transforming in the $G_2$ representations
\begin{equation}
   \rep{7}\times\rep{7}=\rep{1} +\rep{7} +\rep{14} +\rep{27}.
\end{equation}
To see how these relate to the Killing spinor equations, one can use the Killing spinor to define a nowhere vanishing three-form which is invariant under the action of $G_2$
\begin{equation}
\label{eq:def-phi}
   \phi_{abc} = \epsilon^t \gamma_{abc} \gamma^{(7)} \epsilon .
\end{equation}
Then the obstruction to the Levi--Civita having $G_2$ holonomy can be expressed in terms of the intrinsic torsion with (see e.g.~\cite{Kaste:2003zd})
\begin{equation}
\begin{aligned}
   \dd \phi &= w_1 * \phi + w_{7} \wedge \phi + w_{27} ,\\
   \dd * \phi &= \tfrac43 w_7 \wedge * \phi + w_{14} \wedge \phi ,
\end{aligned}
\end{equation}
where $w_i \in \Gs{W_i}$. The derivatives $\dd \phi$ and $\dd * \phi$ contain the same information as $\nabla\phi$, which in turn is fixed by the Killing spinor equations~\eqref{eq:elevendsusy} via~\eqref{eq:def-phi}. We thus have that the fluxes will arrange themselves in terms of these torsion classes, allowing one to classify different solutions. For Minkowski backgrounds, requiring a $G_2$ structure actually implies the flux and warp factor vanish, and so the manifold has $G_2$ holonomy~\cite{Behrndt:2003uq}. However, for compactifications to $AdS_4$, one can consider, for example, the case where only the singlet component $w_1$ is non-vanishing (what is known as a weak $G_2$ manifold) with a nonzero seven-form flux $\tF \propto * w_1$~\cite{Acharya:1998db}. 

\section{Proof of generalised Ricci-flatness}
\label{app:ricci}

We give a brief proof for the $E_{7(7)}\times\bbR^+$ case that a manifold with generalised special holonomy $G$ that stabilises at least one spinor is generalised Ricci-flat. Following the $SU(8)$ conventions of~\cite{CSW3}, using $SU(8)$ indices, and given a generic spinor $\varepsilon$, the scalar and non-scalar components of the generalised Ricci are defined, respectively, by
\begin{equation}
\begin{aligned}
   \tfrac16 \GenS\varepsilon^\alpha 
      &= - \tfrac23 \left( \{ \Dgen^{\alpha \gamma}, \bar\Dgen_{\beta \gamma} \}
            - \tfrac18 \delta^\alpha {}_{\beta} 
               \{ \Dgen^{\gamma \delta}, \bar\Dgen_{\gamma \delta} \}
            \right) \varepsilon^\beta 
      \\ & \qquad \qquad 
         - \tfrac13 \left( 
            [ \Dgen^{\alpha \gamma},  \bar\Dgen_{\beta \gamma} ]
               - \tfrac18 \delta^\alpha {}_{\beta} 
               [ \Dgen^{\gamma \delta},  \bar\Dgen_{\gamma \delta}]
            \right) \varepsilon^\beta
	 - \tfrac18 \BLie{\Dgen^{\beta\gamma}}{\bar\Dgen_{\beta \gamma}}
            \varepsilon^{\alpha} , \\
   \GenRic_{\alpha \beta \gamma \delta} \varepsilon^\delta 
      &= -2 \big( 
         \bar\Dgen_{[\alpha \beta} \bar\Dgen_{\gamma \delta]}
         + \tfrac{1}{4!} \, 
            \epsilon_{\alpha\beta\gamma\delta\epsilon\epsilon'\theta\theta'} 
            \Dgen^{\epsilon \epsilon'} \Dgen^{\theta \theta'} 
            \big)\varepsilon^\delta
         - \BLie{\bar\Dgen_{[\alpha \beta}}{\bar\Dgen_{\gamma] \delta}} 
            \varepsilon^\delta .
\end{aligned}
\end{equation}
If we assume the manifold has generalised special holonomy such that $\varepsilon$ is one of the structure-defining spinors, then we have that $D\varepsilon = 0$ and so $\GenS\varepsilon^\alpha = 0$ and $\GenRic_{\alpha \beta \gamma \delta} \varepsilon^\delta = 0$. We immediately conclude that the scalar part $\GenS = 0$. To see that the remaining components $\GenRic_{\alpha \beta \gamma \delta}$ also vanish, consider that
\begin{equation}
\begin{aligned}
	0 = 9 \epsilon^{[\alpha_1 \dots \alpha_5 \beta_1 \dots \beta_3} 
		\GenRic_{\beta_1 \dots \beta_3 \gamma} \varepsilon^{\gamma]}
	= 5 (4!) (* \GenRic ){}^{[\alpha_1 \dots \alpha_4} \varepsilon^{\alpha_5]} .
\end{aligned}
\end{equation}
The complex self-duality of the generalised Ricci $\GenRic_{\alpha_1\dots\alpha_4} = \tfrac{1}{4!}\epsilon_{\alpha_1\dots\alpha_4\beta_1\dots\beta_4}(\bar\GenRic)^{\beta_1\dots\beta_4}$ then means that we have the equations
\begin{equation}
\begin{aligned}
	(\bar\GenRic)^{\alpha \beta \gamma \delta} \bar\varepsilon_\delta &= 0 ,\\
	 (\bar\GenRic){}^{[\alpha_1 \dots \alpha_4} \varepsilon^{\alpha_5]} &= 0 .
\end{aligned}
\end{equation}
Examining these equations in a basis in which $\varepsilon = (\zeta, 0 , \dots , 0)$ (where $\zeta \bar\zeta = 1$), it becomes clear that they require all components of $\GenRic$ to vanish. Therefore, we have shown that the manifold is generalised Ricci-flat.

For completeness, let us also present an alternative proof that is more analogous to the more common one in conventional geometry. We start with the trivial identity
\begin{equation}
(\bar\GenRic)^{[\alpha_1\dots}(\bar\GenRic)^{\alpha_5\dots}\varepsilon^{\alpha_9]} = 0. 
\end{equation}
Contracting with the $\SU(8)$ invariant epsilon symbol and expanding the antisymmetrisation,
\begin{align}
&\epsilon_{\alpha_1\dots\alpha_8}\left((\bar\GenRic)^{\alpha_1\dots}(\bar\GenRic)^{\alpha_5\dots\alpha_8}\varepsilon^{\alpha_9}
+(\bar\GenRic)^{\alpha_9\alpha_1\dots} (\bar\GenRic)^{\alpha_4\dots}\varepsilon^{\alpha_8} + \dots \right) =\\
&\epsilon_{\alpha_1\dots\alpha_8}\left((\bar\GenRic)^{\alpha_1\dots}(\bar\GenRic)^{\alpha_5\dots\alpha_8}\varepsilon^{\alpha_9} + 8 (\bar\GenRic)^{\alpha_9\alpha_2\alpha_3\alpha_4} (\bar\GenRic)^{\alpha_5\dots\alpha_8}\varepsilon^{\alpha_1}\right) = 0 .
\end{align}
Then using the complex self-duality of $\GenRic$, we obtain
\begin{equation}
\left(\GenRic_{\alpha_1\dots\alpha_4}(\bar\GenRic)^{\alpha_1\dots\alpha_4}\right)\varepsilon^{\alpha_9} + 8 (\bar\GenRic)^{\alpha_9\alpha_2\alpha_3\alpha_4} \GenRic_{\alpha_1\dots\alpha_4}\varepsilon^{\alpha_1} = 0,
\end{equation}
but the second term in the left-hand side vanishes due to the generalised special holonomy, so we conclude
\begin{equation}
\GenRic_{\alpha_1\dots\alpha_4}(\bar\GenRic)^{\alpha_1\dots\alpha_4} = 0.
\end{equation}
This is positive-definite so it implies that $\GenRic = 0$ and again we conclude the full generalised Ricci vanishes. In other dimensions we will generically obtain equivalent expressions $\GenS = 0$ and $G(\GenRic,\GenRic)=0$, where $G$ is the generalised metric which is positive-definite in manifolds with Euclidean signature, thus again implying generalised Ricci-flatness.

%%%%%%%%%%%%%%%%%%%%%%%

\section{Detailed construction of $Q$}
\label{app:3M5}

Here we show how to construct $\slashed{Q}$ such that $\Dgen \varepsilon_0 = 0$ as claimed in section~\ref{sec:3M5}. Substituting the expression~\eqref{eq:3M5-flux} for the flux into equation~\eqref{eq:Dgen-1} we see that we require
\begin{equation}
\label{eq:Q1}
\begin{aligned}
	\slashed{Q}_{u^i} \varepsilon_0 &= - \ee^{2\Delta} X_i\Gamma_{u^i} \varepsilon_0 \\
	&\text{where} \hspace{20pt}
	X_i = - \tfrac14 \der \Delta - \tfrac18 \der f_i ,
\end{aligned}
\end{equation}
together with $\slashed{Q}_z \varepsilon_0 = 0$. It is easy to see that taking
\begin{equation}
\label{eq:Q1-expression}
\begin{aligned}
	Q_{u^i z v^i} &= - Q_{zu^i v^i} = - 2 X_i g_{u^i v^i}   ,
\end{aligned}
\end{equation}
with all remaining components vanishing, as in equation~\eqref{eq:Qmn}, this satisfies~\eqref{eq:Q1}. An important feature of the quantities $X_i$ is that
\begin{equation}
	X_1 + X_2 + X_3 = 0 .
\end{equation}
Using this relation, one can see that the defined $Q_{pqm}$ is indeed traceless, and so transforms in the $\rep{105}$ representation of $\SO(7)$.

Similarly, from~\eqref{eq:Dgen-2} we find
\begin{equation}
\label{eq:Q2}
\begin{aligned}
	\slashed{Q}^{u^i v^j} \varepsilon_0 
	&= -\ee^{2\Delta} A_{ij} \Gamma^{u^i v^j} \varepsilon_0 ,\\
	\slashed{Q}^{u^i z} \varepsilon_0 &= 0 ,
\end{aligned}
\end{equation}
where
\begin{equation}
\begin{aligned}
	A_{ij} = \tfrac3{20} \der \Delta + \tfrac1{24}( \der f_i + \der f_j ) - \tfrac1{24} \delta_{ij}  (\der f_i) .
\end{aligned}
\end{equation}
We use the ansatz from~\eqref{eq:3M5-Q}
\begin{equation}
\begin{aligned}
	\slashed{Q}^{mn} &= Q^{mn}{}_{pq} \Gamma^{pq}
		+ \tfrac18   Q_{pq}{}^{[m} \Gamma^{n]pq} ,
\end{aligned}
\end{equation}
with $Q_{mnp}$ as in~\eqref{eq:Q1-expression} and
\begin{equation}
\begin{aligned}
	Q^{u^i v^j}{}_{u'^i v'^j} 
		&= \ee^{2\Delta} Y_{ij} \delta^{u^i v^j}_{u'^i v'^j}  
		\hspace{20pt} \text{where} \hspace{10pt}
		Y_{ij} = - ( A_{ij} + \tfrac14 (X_i + X_j) )	,
\end{aligned}
\end{equation}
and all remaining components are zero. As $Y_{ij}$ satisfies the relations
\begin{equation}
\label{eq:trace}
\begin{aligned}
	\sum_j Y_{ij} = \tfrac12 Y_{ii} ,
\end{aligned}
\end{equation}
we see that $Q^{mn}{}_{pq}$ is traceless and hence transforms in the $\rep{168}$ representation. This then solves~\eqref{eq:Q2}.

Finally,~\eqref{eq:Dgen-3} and~\eqref{eq:Dgen-4} imply simply
\begin{equation}
\label{eq:Q571}
	\slashed{Q}^{m_1 \dots m_5} \varepsilon_0 = 0
	\hspace{25pt} \text{and} \hspace{25pt}
	\slashed{Q}^{m,m_1 \dots m_7} \varepsilon_0 = 0 ,
\end{equation}
which we can solve trivially by taking
\begin{equation}
\begin{aligned}
   \slashed{Q}^{m_1 \dots m_5}  &= 0 ,\\
   \slashed{Q}^{m,m_1 \dots m_7} &= 0 ,
\end{aligned}
\end{equation}
since the torsion map does not mix these components with the ones we already fixed.

We have therefore constructed a tensor $Q$ which completes the generalised connection~\eqref{eq:Dgen-Q} so that $\Dgen \varepsilon_0 = 0$.

%%%%%%%%%%%%%%%%%%%%%%%%%%%%%%%%%%%%%%%%%%%%%%%%%%%%%%%%%%%%%%%%%%%%%%%%%%

%%%%%%%%%%%%%%%%%%%%%%%%%%%%%%%%%%%%%%%%%%%%%%%%%%%%%%%%%%%%%%%%%%%%%%%%%%

\end{document}